\documentclass[iop]{emulateapj}
\pdfoutput=1 
\usepackage{graphicx}
\usepackage{amsmath}

\usepackage{hyperref}
\usepackage{color}

\newcommand{\flag}[1]{\textsc{\small{#1}}}
\newcommand{\am}[0]{[$\alpha$/M]}
\newcommand{\mh}[0]{[M/H]}
\newcommand{\xh}[1]{[#1/H]}
\newcommand{\xm}[1]{[#1/M]}
\newcommand{\xfe}[0]{[X/Fe]}
\newcommand{\feh}[0]{[Fe/H]}
\newcommand{\cm}[0]{[C/M]}
\newcommand{\nm}[0]{[N/M]}
\newcommand{\teff}[0]{$T_{\rm eff}$}
\newcommand{\logg}[0]{$\log{g}$}

\begin{document}

\title{Abundances, stellar parameters, and spectra from the SDSS-III/APOGEE Survey}

\author{
Jon A. Holtzman\altaffilmark{1},
Matthew Shetrone \altaffilmark{2},
Jennifer A. Johnson\altaffilmark{3},
Carlos Allende Prieto\altaffilmark{4,5},
Friedrich Anders\altaffilmark{6,7},
Brett Andrews\altaffilmark{8},
Timothy C. Beers\altaffilmark{9},
Dmitry Bizyaev \altaffilmark{10},
Michael R. Blanton\altaffilmark{11},
Jo Bovy\altaffilmark{12,13},
Ricardo Carrera\altaffilmark{4},
Katia Cunha\altaffilmark{14,15},
Daniel J. Eisenstein\altaffilmark{17},
Diane Feuillet\altaffilmark{1},
Peter M. Frinchaboy\altaffilmark{17},
Jessica Galbraith-Frew\altaffilmark{18},
Ana E. Garc\'{\i}a P\'erez\altaffilmark{4},
D. Anibal Garc\'{\i}a Hern\'andez\altaffilmark{4},
Sten Hasselquist\altaffilmark{1},
Michael R. Hayden\altaffilmark{1},
Fred R. Hearty\altaffilmark{19},
Inese Ivans\altaffilmark{18},
Steven R. Majewski\altaffilmark{20},
Sarah Martell\altaffilmark{21},
Szabolcs Meszaros\altaffilmark{22},
Demitri Muna\altaffilmark{3},
David Nidever\altaffilmark{23},
Duy Cuong Nguyen\altaffilmark{24},
Robert W. O'Connell\altaffilmark{20},
Kaike Pan\altaffilmark{10},
Marc Pinsonneault\altaffilmark{3},
Annie C. Robin\altaffilmark{25},
Ricardo P. Schiavon\altaffilmark{26},
Neville Shane\altaffilmark{20},
Jennifer Sobeck\altaffilmark{20},
Verne V. Smith\altaffilmark{14},
Nicholas Troup\altaffilmark{20},
David H. Weinberg\altaffilmark{3},
John C. Wilson\altaffilmark{20},
W. M. Wood-Vasey\altaffilmark{8},
Olga Zamora\altaffilmark{4},
Gail Zasowski\altaffilmark{27}
}

\altaffiltext{1}{New Mexico State University, Las Cruces, NM 88003, USA (holtz@nmsu.edu, mrhayden@nmsu.edu, feuilldk@nmsu.edu)}

\altaffiltext{2}{University of Texas at Austin, McDonald Observatory, Fort Davis, TX 79734, USA (shetrone@astro.as.utexas.edu)}

\altaffiltext{3}{Department of Astronomy, The Ohio State University, Columbus, OH 43210, USA (jaj@astronomy.ohio-state.edu, muna@astronomy.ohio-state.edu)}

\altaffiltext{4}{Instituto de Astrof\'{\i}sica de Canarias, 38205 La Laguna, Tenerife, Spain}

\altaffiltext{5}{Departamento de Astrof\'{\i}sica, Universidad de La Laguna,
38206 La Laguna, Tenerife, Spain (callende@iac.es, agp@iac.es)}

\altaffiltext{6}{Leibniz-Institut fur Astrophysik Potsdam (AIP), An der Sternwarte 16, 14482, Potsdam, Germany}

\altaffiltext{7}{Laborat\'{o}rio Interinstitucional de e-Astronom\'{i}a (LIneA), Rua Gal. Jos\'{e} Cristino 77, Rio de Jane\'{i}ro, RJ - 20921400, Brasil}

\altaffiltext{8}{PITT PACC, Department of Physics and Astronomy, University of Pittsburgh, Pittsburgh PA 15260, USA, wmwv@pitt.edu }

\altaffiltext{9}{Dept. of Physics and JINA Center for Evolution of the Elements, University of Notre Dame, Notre Dame, IN 46556 USA}

\altaffiltext{10}{Apache Point Observatory, P.O. Box 59, Sunspot, NM 88349-0059, USA (dmbiz@apo.nmsu.edu)}

\altaffiltext{11}{ Center for Cosmology and Particle Physics, Department of Physics, New York University, 4 Washington Place, New York, NY 10003, USA, michael.blanton@gmail.com}

\altaffiltext{12}{\label{1}Institute for Advanced Study, Einstein Drive, Princeton, NJ 08540, USA, bovy@ias.edu}

\altaffiltext{13}{\label{2}John Bahcall Fellow}

\altaffiltext{14}{National Optical Astronomy Observatories, Tucson, AZ 85719, USA (vsmith@email.noao.edu, cunha@email.noao.edu)}

\altaffiltext{15}{Observat\'orio Nacional, S\~ao Crist\'ov\~ao, Rio de Janeiro, Brazil}

\altaffiltext{16}{Harvard-Smithsonian Center for Astrophysics, 60 Garden St., Cambridge, MA 02138, deisenstein@cfa.harvard.edu}

\altaffiltext{17}{Texas Christian University, Fort Worth, TX 76129, USA (p.frinchaboy@tcu.edu)}

\altaffiltext{18}{Physics and Astronomy, University of Utah, Salt Lake City, UT 84112, USA (iii@physics.utah.edu)}

\altaffiltext{19}{Department of Astronomy and Astrophysics, The Pennsylvania State University, University Park, PA 16802; Institute for Gravitation and the Cosmos, The Pennsylvania State University, University Park, PA 16802, USA (frh10@psu.edu)}

\altaffiltext{20}{Dept. of Astronomy, University of Virginia,
Charlottesville, VA 22904-4325, USA (srm4n, aeg4x, rwo@virginia.edu)}

\altaffiltext{21}{School of Physics, University of New South Wales, Sydney NSW 2
052}

\altaffiltext{22}{ELTE Gothard Astrophysical Observatory, H-9704 Szombathely, Szent Imre herceg st. 112, Hungary}

\altaffiltext{23}{Department of Astronomy, University of Michigan, Ann Arbor, MI 48109, USA (dnidever@umich.edu)}

\altaffiltext{24}{Dunlap Institute for Astronomy and Astrophysics, University of Toronto, Toronto, Ontario, Canada}

\altaffiltext{25}{ Institut UTINAM, OSU THETA, University of Franche-Comt\'e, Besan\c{c}on, France (annie@obs-besancon.fr)}

\altaffiltext{26}{Astrophysics Research Institute, Liverpool John Moores University, Liverpool, L3 5RF, United Kingdom (rpschiavon@gmail.com)}

\altaffiltext{27}{Department of Physics and Astronomy, Johns Hopkins University, Baltimore, MD 21218, USA (gail.zasowski@gmail.com)}

%
%
%
%

\begin{abstract}
The SDSS-III/APOGEE survey operated from 2011-2014 using the APOGEE
spectrograph, which collects high-resolution ($R\sim 22,500$), near-IR
($1.51-1.70\mu$m) spectra with a multiplexing (300 fiber-fed objects)
capability. We describe the survey data products that are publicly
available, which include catalogs with radial velocity, stellar
parameters, and 15 elemental abundances for over 150,000 stars,
as well as the more than 500,000 spectra from which these quantities
are derived. Calibration relations for the stellar parameters (\teff,
\logg, \xh{M}, \am) and abundances (C, N, O, Na, Mg, Al, Si, S, K, Ca,
Ti, V, Mn, Fe, Ni) are presented and discussed.  The internal scatter
of the abundances within clusters indicates that abundance precision is
generally between 0.05 and 0.09 dex across a broad temperature range;
within more limited ranges and at high S/N, it is smaller for some elemental
abundances. We assess the accuracy of the abundances using comparison of
mean cluster metallicities with literature values, APOGEE observations of
the solar spectrum and of Arcturus, comparison of individual star
abundances with other measurements, and consideration of the locus of
derived parameters and abundances of the entire sample, and find that
it is challenging to determine the absolute abundance scale; external
accuracy may be good to 0.1-0.2 dex. Uncertainties may be larger at
cooler temperatures (\teff $<4000$K). Access to the public data
release and data products is described, and some guidance for using the
data products is provided.

\end{abstract}

\section{Introduction}

The third phase of the Sloan Digital Sky Survey (SDSS-III) included 
the Apache Point Observatory Galactic Evolution Experiment (APOGEE) as
one of its key components. APOGEE made use of a new, multi-object,
near-IR spectrograph to obtain spectra of more than 150,000 stars across
portions of the Milky Way visible from the Apache Point Observatory
(APO).  The main goal of the APOGEE survey is to obtain high-resolution
spectra to map out the kinematical and chemical structure of the Milky Way.

The overall design and goals of the SDSS-III survey are described in
\citet{Eisenstein2011} and of APOGEE in 
\citet{Majewski2015}.  The survey had a period of commissioning
in spring 2011 and operated in survey-mode from fall 2011 until
summer 2014. Data from the first year of operation was released as
part of the SDSS DR10 \citep{DR10} and included measurements of
basic stellar parameters as well as overall metal and $\alpha$-element
abundances.  An SDSS-III internal data release (DR11) included the first two
years of survey data with the same analysis as was used for DR10.
The final SDSS-III data release (DR12; \citealt{DR12})
includes the full 3 years of data from the SDSS-III/APOGEE survey.
For DR12, all of the data have been reprocessed, and the parameter
and abundance analysis have been significantly updated: APOGEE DR12
includes measurements of individual chemical abundances for 15
different elements.

The operation of the APOGEE instrument will continue in the APOGEE-2
survey, which is part of the fourth phase of the Sloan Digital Sky
Survey (SDSS-IV) that plans to operate from 2014 to 2020.  APOGEE-2
will consist of additional observations in the northern hemisphere
from APO but will also include observations of the southern Milky
Way that will be obtained with the 2.5m du Pont telescope at Las Campanas 
Observatory. Future data releases for this survey may include reprocessing
of APOGEE-1 data.

This paper provides a general overview of the data provided by the
SDSS-III/APOGEE survey. \S \ref{sect:survey} briefly summarizes the
survey. \S \ref{sect:reduction} and \S \ref{sect:spectro} summarize
how the reduction and spectroscopic analysis is performed 
(but described in greater detail in \citealt{Nidever2015} and
\citealt{GarciaPerez2015}) and introduce the reader
to the associated data products. \S \ref{sect:calibration} discusses
the calibration and validation of the derived stellar parameters and
abundances. Options for accessing the data are presented in \S
\ref{sect:access}. \S \ref{sect:usingdata} describes the
APOGEE catalog data and presents some important considerations for users of 
these data, and \S \ref{sect:usingspectra} has a similar presentation
for the spectra.

\section{The SDSS-III/APOGEE survey}
\label{sect:survey}

\subsection{The APOGEE instrument}

The APOGEE survey utilizes a fiber spectrograph that records 300
spectra simultaneously. The spectrograph uses a volume phase
holographic (VPH) grating and camera optics to deliver spectra in
the H band between 1.51 and 1.70  $\mu$m  at a spectral resolution
of $R\sim22,500$. The spectra are recorded on three Hawaii-2 RG
detectors with small gaps in spectral coverage between the detectors.
The instrument was constructed largely at the University of Virginia
and is described in detail in \citet{Wilson2015}.

The APOGEE spectrograph was shipped to Apache Point Observatory in
the spring of 2011 and was commissioned during the period April-June
of 2011.  During the commissioning period, it was recognized that
there was some astigmatism within the instrument, so modifications
were made during summer 2011 to address this.  Because of the optical
issues and modifications after commissioning, the commissioning data
have lower spectral resolution that does not meet the strict survey
requirements, especially in the detector covering the longest
wavelengths.  Commissioning data have been released in the APOGEE
data releases as these data provide useful radial velocity information,
but abundance analysis is not currently provided for them because
of the reduced resolution and quality. However, a number of the
sources observed during commissioning were reobserved during the
main survey.

After the summer 2011 modifications, the instrument was pumped down and
cooled in August 2011 and remained unopened, operating continuously
without any large changes
in pressure or temperature, for the duration of the SDSS-III/APOGEE survey.
Over the period of the main survey, the instrument
met survey requirements and was remarkably stable, with essentially
no changes in optical and detector performance.  

At the shorter end of the wavelength range, the detector pixels do
not quite adequately sample the spectrograph point-source function.
To ensure adequate spectral sampling, the instrument includes a dither mechanism
for moving the entire detector array, and exposures are taken in
pairs, with the detectors shifted 0.5 pixels between the two
observations in a pair.

One of the detectors (at the short wavelength end) used in the
SDSS-III/APOGEE survey exhibits strong persistence, where a significant
fraction of accumulated charge is released over a long period of
time \citep{Nidever2015}. This affects roughly a third of the chip
and is distributed in such a way that about a third of the fibers are
affected at all wavelengths on that chip (1.51$\mu$m-1.58$\mu$m), while the other
two-thirds of the fibers are unaffected. The behavior of the persistence
is challenging to characterize, and its relevance depends both on the
brightness of a spectrum as well as the history of previous exposures.
The current reduction analysis does not attempt to correct for this; we
present some information on the impact on derived stellar parameters and
abundances in \S \ref{sect:persist}. The offending detector was swapped
out in summer 2014 after the completion of the SDSS-III/APOGEE survey.

\subsection{Survey operations}

The primary source of the light to the spectrograph is the SDSS 2.5m
telescope \citep{Gunn2006}, which uses a plug-plate system to collect
light from a circular 3-degree diameter field. For the vast majority of the
observations, 230 of the fibers were placed on science targets,  35
fibers on hot star targets used for correction of telluric absorption
(although these also enable some science projects), and 35 fibers on
``blank" sky regions to enable sky subtraction.

Standard exposure sequences consist of 8 separate 500s exposures, with
4 at each of the two dither positions, in an ABBAABBA pattern.  The
4000s integration time yields a S/N of about 100 per half-resolution
element (a typical pixel) on a source with $H=11$, although
the sensitivity varies across
the spectral region recorded on the detectors, primarily as a result
of grating sensitivity changes.

\subsection{1m telescope fiber feed}

Partway through the survey, a fiber feed was constructed 
from the nearby NMSU 1m telescope for 10
fibers, which are located in a fixed linear pattern in the focal
plane of that telescope.  This configuration enables observations
of single objects, with accompanying sky spectra, and has been used
to provide a sample of mostly bright stars, useful for both calibration
and science, as discussed in \S \ref{sect:refcal}. To 
use the 1m observations to validate those obtained with the 2.5m,
the spectra from the two telescopes must match closely. The input beam from the
2.5m telescope is f/5, while that of the 1m telescope is f/6. However,
the fiber run from the 1m telescope adds significant fiber to the
standard 2.5m configuration and includes
several additional couplings, so the focal ratio degradation is expected
to be a bit worse for the 1m feed, which goes in the direction of making
the beam more closely match that obtained with the 2.5m.

\begin{figure}[h]
\includegraphics[width=0.5 \textwidth]{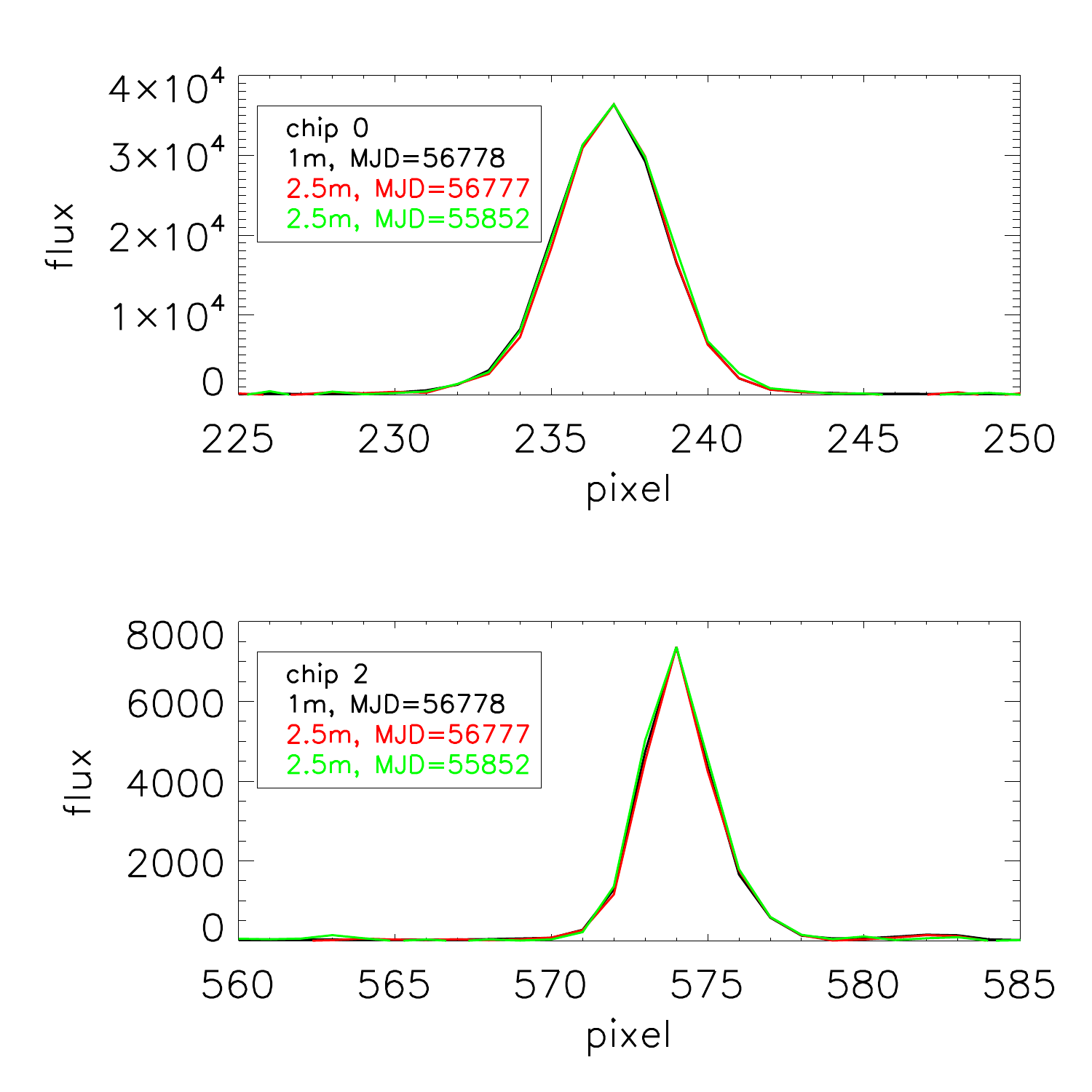}
\caption{Comparison of sky emission lines as taken with the 2.5m and
1m telescopes.
}
\label{fig:sky}
\end{figure}

Figure \ref{fig:sky}
shows the profile of two bright sky lines, one in the long wavelength
chip (chip 0) and one in the short wavelength chip (chip 2), for several
1m and 2.5m observations. This demonstrates that the line profiles
as obtained with APOGEE from the two telescopes match extremely
well. Comparison of spectra obtained for the same star from both
telescopes also shows the same good match, although the different S/N
of the spectra make detailed comparison difficult. Both sky and stellar spectra
suggest that the NMSU 1m delivers extremely similar spectra to those
delivered by the 2.5m telescope.

\subsection{Target selection}

Target selection is described in detail in \citet{Zasowski2013}.
The ``main" sample is selected using a simple color cut with
$(J-K)_0>0.5$, where colors are dereddened using a combination of
near-IR and mid-IR photometry \citep{Majewski2011}, but there are
also other target classes, including calibration and ancillary
science program targets, as discussed in more detail in
\citet{Zasowski2013}.

For most of the stars, data are collected in multiple visits to
allow detection of radial velocity variables. Standard survey
targets are observed in multiple visits until a $S/N$ of 100 per
half-resolution element is achieved for a target of H=12.2.  A
minimum of three visits is required for most targets, with restrictions
on the cadence so that a broad range of RV variables can be detected.
However, survey software tracks accumulated $S/N$, and additional
visits are added as needed to achieve the $S/N$ requirement. Some
fainter targets (medium and long cohorts; see \citealt{Zasowski2013})
were observed for more than three visits.  A subset of survey targets
towards the Galactic Bulge was observed only in single visits (with
a brighter limit for sample selection), because of the challenge
associated with observing such targets from Apache Point Observatory,
where it is always at a high airmass.

\subsection{Scope of APOGEE data releases}

The APOGEE data in DR10 \citep{DR10} included the first year of
APOGEE observations. The SDSS/APOGEE DR11 release added the second
year of data, but was initially only released internally to SDSS
collaboration members; it is included in the public DR12.  Calibration
of the DR10 and DR11 data is described in \citet{Meszaros2013}.

The SDSS DR12 \citep{DR12} includes all data taken with the APOGEE
instrument between April 2011 and July 2014. A total of 163,281 stars are
included. Of these, 151,141 stars were observed during the main survey
period that provided survey-quality spectra. Hot stars for telluric
correction account for 15,343 of the survey-quality targets, and 17,116 of
all targets (including commissioning data).  Of the 135,798 survey-quality
non-telluric sources, 110,087 come from the ``main-survey" sample, which does
not include special targets, targets from ancillary science programs,
and stars observed with the NMSU 1m (883 stars).

\section{Basic data reduction and associated data products}
\label{sect:reduction}

The basic data reduction procedure is described in \citet{Nidever2015}. 
We provide a brief synopsis as an introduction to the data products
that are included in the data releases and described here.

The raw data consist of data cubes for each individual exposure, where
each layer of the cube is a read of the three detectors every 10.7 seconds.
As a result, the cubes are large, with a standard 500s exposure consisting
of 47 reads of 3 detectors, with each detector having 2560x2048 pixels,
since a ``reference'' array (intended to help with tracking bias variations)
of 512x2048 pixels is included. The raw
data cubes are compressed using the standard FPACK routine \citep{seaman2010}, but,
to provide maximum compression, the raw data are converted to a series of
differences between adjacent reads before compression. The resulting
files are saved as \textit{apR} files.

The reduction procedure reduces these image cubes to individual 2D
images. For each read, dark current is subtracted using a calibration
product constructed from multiple dark images.  Up-the-ramp sampling is
then used to derive count rates that are less affected by readout noise 
than a simple doubly-correlated sampling would provide. The resulting
image is flat fielded using a calibration product constructed from
multiple flat-field images and is stored in an \textit{ap2D} file.

Spectra are optimally extracted from the 2D images using an empirical
PSF derived from an exposure of illuminated telescope mirror covers
taken after the science exposure sequence has been completed. An
approximate relative flux correction is also applied. These extracted,
flux-calibrated spectra are saved as \textit{ap1D} files, and the
2D models used for the extraction are saved as \textit{ap2Dmodel}
files.

The multiple exposures (\textit{ap1D} files) are then combined to
make an individual visit spectrum. To do this, the relative shifts
between the different exposures, arising from the motion of the
detector dither mechanism used to provide better sampling, are
determined. Position-dependent sky subtraction is performed on
individual exposures, as is a correction for telluric absorption.
Exposures at different dither positions are paired, matching as
best as possible exposures of comparable S/N, and combined into a
spectrum with a sampling corresponding to half the original pixel
sampling. The combination technique reduces to simple interleaving
(after adjusting the levels of the two exposures to match) in the
case of perfect 0.5 pixel dithers but properly handles dithers
that are imperfect. The combined pairs are then coadded to provide
the final visit spectra. An approximate absolute flux calibration
is made to the spectrum using the 2MASS $H$ magnitude.  An initial
measurement of the radial velocity for each star in the visit is
made by cross correlating each spectrum with the best match in a
template library, and these are stored against the visit spectra in
\textit{apVisit} files.

All of the visits are then combined after resampling them to a common,
log-lambda wavelength scale; in all data products, the wavelength scale
is in vacuum wavelengths. Relative radial velocities of each visit are
iteratively refined by cross-correlation of each visit spectrum with the
combined spectrum (which should provide a perfect template match). The absolute
radial velocity scale is then set by cross-correlating the combined spectrum
with the best match in a template library. The resulting combined spectrum,
as well as the resampled visit spectra, are stored in \textit{apStar} files.

\begin{deluxetable*}{ll}[t]
\tablecaption{APOGEE Spectral data products}
\tablehead{
\colhead{Name}&\colhead{Contents}
}
\startdata
\textit{apR-[abc]-ID8.apz} & Raw data cubes (2560x2048x47), APOGEE compressed format \\
\textit{ap2D-[abc]-ID8.fits} & 2D (2048x2048) dark subtracted, flat fielded\\
\textit{ap1D-[abc]-ID8.fits} & 1D extracted (2048x300), rough wavelength calibration (not zero-pointed using sky)\\
\textit{apCframe-[abc]-ID8.fits} & 1D spectra with sky subtraction, telluric correction, wavelength zeropoint adjustment\\
\textit{apVisit-APRED\_VERS-PLATE-MJD-FIBER.fits} & Individual visit spectra, dither-combined, but otherwise native pixel scale, \\
& one per object visit\\
\textit{apStar-APRED\_VERS-APOGEE\_ID.fits} & Spectra resampled to constant $\Delta\log \lambda$, along with combined spectrum, one per object\\
\textit{aspcapStar-RESULTS\_VERS-APOGEE\_ID.fits} & Normalized spectra to match
those of spectral grids, \\
& with best-matching grid spectra, one per object \\
\enddata
\label{tab:data}
\end{deluxetable*}

A summary of the spectral data products is presented in Table \ref{tab:data}.
The DR12 Science Archive Server (\url{http://data.sdss3.org/sas/dr12})
makes both raw data, as well as various stages of reduced data,
available to the community, as discussed in \S \ref{sect:access}.
More details on the files are presented in \S \ref{sect:usingspectra}.

The reduction pipeline software is version-controlled in the SDSS-III 
software repository. 

\section{Spectroscopic analysis of APOGEE data and associated data catalogs}
\label{sect:spectro}

Stellar parameters and individual chemical abundances are derived from the
combined APOGEE spectra with the APOGEE Stellar Parameters and Chemical
Abundances Pipeline (ASPCAP), which is described in detail in 
\citet{GarciaPerez2015}. The basic idea is that a grid of synthetic
spectra is searched (with interpolation in the grid) to find the best
match to each observed spectrum, and the parameters of the best match
are adopted as the best parameters for that star.

\subsection{Synthetic spectra}

Central to ASPCAP is the construction of a synthetic spectral grid that
covers the range of fundamental stellar parameters over which main
survey stars are expected to be found.  Given the significance and
frequency of molecular features --- in particular, from CO, CN, and OH
--- in the region of the $H$-band where APOGEE spectra are recorded, the
dimensions of this grid include not only the usual effective temperature (\teff),
surface gravity (\logg), microturbulent velocity ($v_{micro}$), 
and overall scaled-solar chemical abundance (\mh), but also the carbon (\cm), 
nitrogen (\nm), and $\alpha$-element abundances (\am, which includes oxygen) 
relative to the solar abundance ratio. The resulting
7D grid comprises several million individual synthetic spectra.

The fidelity of the synthetic spectral library is critical to the 
success of the method. The structure of the underlying stellar atmospheres 
were computed with the ATLAS9 code \citep{Kurucz1979}. For DR10/DR11, a grid
of atmospheres was used that adopted solar-scaled abundances  (i.e.,
the carbon, nitrogren, and $\alpha$ abundances were varied only in the
spectral synthesis, but not in the underlying atmospheric model structure).
For DR12, a self-consistent set of atmospheres were used: 
carbon and $\alpha$ abundances were varied across the
grid for the atmosphere calculation (the nitrogen abundance has
little effect on the structure of the atmosphere, so this was not
varied).  Details are given in \citet{Meszaros2012}. These calculations
assume plane-parallel atmospheres (but see \citealt{Zamora2015} for some
discussion of the changes resulting from using a spherical geometry) 
in local thermodynamic equilibrium (LTE).

The synthetic spectra were generated using the LTE-1D branch of
the ASS$\epsilon$T synthesis code \citep{Koesterke2009}.  A custom
line list was compiled as discussed in \citet{Shetrone2015}. The
atomic and molecular data for the line list were taken from various
literature sources, but the atomic data transition probabilities
were adjusted, within limits, by matching synthetic spectra for
the Sun and Arcturus with available observed high-resolution
atlases \citep{Livingston1991,Hinkle1995} as described in
\citet{Shetrone2015}. The solar abundances assumed for these adjustments
are those from \citet{Asplund2009} and \citet{Asplund2005} (for DR10/11
and DR12, respectively) and those for Arcturus from consideration
of a number of sources; the adopted values for the various APOGEE
line lists are presented in \citet{Shetrone2015}.  As described in
\citet{Shetrone2015}, the linelist has evolved, and future development
is likely.  Additional details on the spectral grids and a comparison
of the ATLAS9/ASS$\epsilon$T synthetic spectra with those from other
atmospheres and syntheses are discussed in \citet{Zamora2015}.


The high-resolution synthetic spectra are then matched to the APOGEE
resolution and pixel sampling. This is accomplished by smoothing the
high-resolution spectrum with the line-spread function (LSF) of APOGEE data. 
However,
the APOGEE LSF varies across the detector as a consequence of the
optical design. To some extent, this variation is reduced when
multiple visits are combined, since stars will generally land in a
different location on the detector (from a different fiber in the telescope 
focal
plane) in different visits. In DR10, a Gaussian LSF with $R=22500$ was
adopted, but we subsequently recognized that the APOGEE LSF is not 
well-represented by a Gaussian. For DR12 we have taken a approach
of using an LSF made by averaging five different empirically-measured LSFs
across the detector to make a single library for use with all of
the SDSS-III/APOGEE spectra. As demonstrated below, this yields results
that are validated by observations of objects of known abundances,
but it is an area for potential improvement in subsequent
data releases.  \citet{GarciaPerez2015} present
a more detailed investigation of the potential impact of assuming
an imperfect LSF.

In addition to smoothing the high-resolution synthetic spectra with
an APOGEE LSF, in DR12 we also include broadening to account for macroturbulent
velocity in the stellar atmosphere.  We were motivated to do this not
only because it is a known physical phenomenon, but also
because this additional broadening brings our derived surface gravities
closer to those suggested by other techniques (see below). For DR12,
we have adopted a simple scheme of broadening with a Gaussian of
FWHM 6 km/s for the macroturbulent velocity, motivated by typical
values from \citet{Carney2008}. Adopting a different
functional form, along with possible variations in macroturbulent
velocity as a function of other stellar parameters, is another area
for potential improvement in the future.

The synthetic spectra are calculated relative to a true continuum.
However, it is challenging, if not impossible, to identify a true
continuum in the observed spectra in cool, metal-rich stars. 
To account for this, \textit{both}
the synthetic spectra and the observed spectra are ``pseudo-continuum''
normalized using the same algorithm, as discussed in 
\citet{GarciaPerez2015}.

The smoothed, resampled, and pseudo-continuum normalized spectra are bundled
into a large library.  To minimize the (substantial) computing
resources required
for such a large grid, we separate the grid into two grids overlapping
in temperature (which we denote GK and F); we find the best match in each grid, and use the result
that provides the better match. In addition, the libraries are compressed
using a principal components decomposition, in which we split the
spectra into 30 pieces and PCA-compress each of those, keeping only the
first 30 principal components and the amplitudes of these for each
location in the grid; additional details are provided in 
\citet{GarciaPerez2015}.

The range and spacing of the ATLAS9/ASS$\epsilon$T grid parameters used for DR12 are 
given in Table \ref{tab:grid} (see also \citealt{Zamora2015}). 
\citet{GarciaPerez2015} discuss why
this grid spacing is adequate and how interpolation within the grid
is performed.

\begin{deluxetable}{llll}
\tablewidth{0pt}
\tablecaption{Synthetic spectral grid parameter values}
\tablehead{
\colhead{gridname} & \colhead{Dimension} & \colhead{Range} & \colhead{Step size}
}
\startdata
GK grid & \teff            & 3500 to 6500     & 250 \\
   & \logg         &  0 to 5          & 0.5 \\
   & $\log v_{micro}$ & -0.301 to 0.903  & 0.301 \\
   & \mh              & -2.5 to 0.5      & 0.5 \\
   & \cm              & -1  to 1         & 0.25\\
   & \nm              & -1  to 1         & 0.5\\
   & \am              & -1  to 1         & 0.25\\
F grid & \teff            & 5500 to 8000     & 250 \\
   & \logg         &  0 to 5          & 0.5 \\
   & $\log v_{micro}$ & -0.301 to 0.903  & 0.301 \\
   & \mh              & -2.5 to 0.5      & 0.5 \\
   & \cm              & -1  to 1         & 0.25\\
   & \nm              & -1  to 1         & 0.5\\
   & \am              & -1  to 1         & 0.25\\
\enddata
\label{tab:grid}
\end{deluxetable}

We note that for SDSS-III/APOGEE, stellar rotation is not included in the
synthetic spectral grids. For stars with rotational velocities ($\geq 
5$ km/s) that broaden the lines significantly beyond the instrument resolution,
the libraries will not be able to obtain a good match to the observed
spectra, and the resulting
stellar parameters can be affected systematically. Stellar rotation
is not expected to be significant for most of the red giants that make
up the APOGEE sample.  However, a significant fraction of the dwarfs
observed in the survey may have detectable rotation, and results for
these are suspect. As discussed below, we do not provide calibrated
parameters for dwarfs, partly because of the lack of treatment of
rotation. For all survey stars, we provide a flag that indicates
whether rotation is likely to be important, based on a comparison of
the width of the autocorrelation function of the spectrum relative to
the autocorrelation of the best-matching RV template.  Future improvements
are planned with the inclusion of rotation when needed.


We also note that the current analysis is limited to stars
with \teff$>3500$ K because the Kurucz grid of model
atmospheres does not extend to
cooler temperatures. Subsequent work is planned to extend results to
cooler temperatures using MARCS model atmospheres and the Turbospectrum
synthesis code \citep{Zamora2015}.

\subsection{Deriving stellar parameters and abundances}

\begin{figure}
\includegraphics[width=0.5 \textwidth]{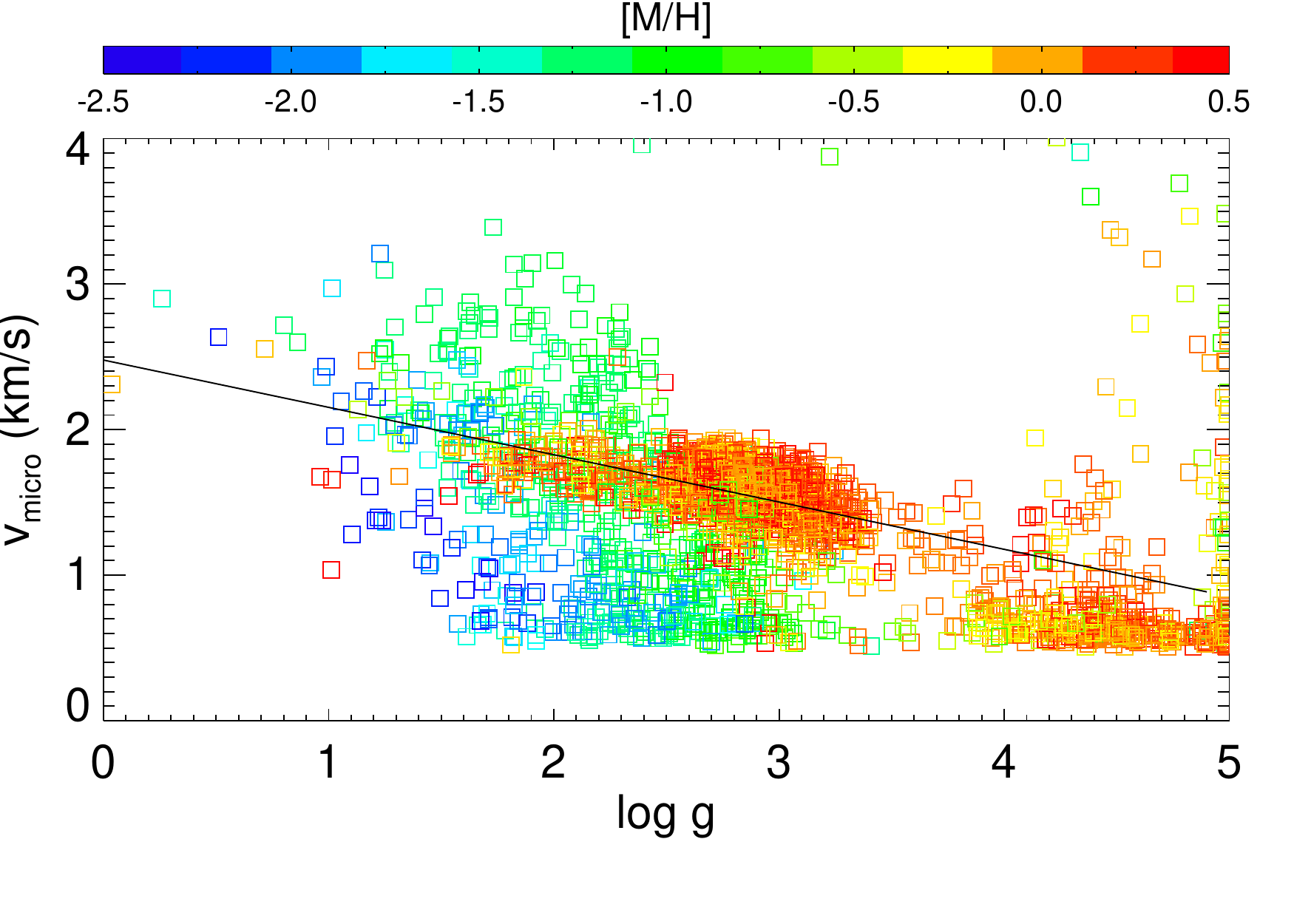}
\caption{Microturbulence relation, as derived from a 7D run of
the calibration subsample.
}
\label{fig:vmicro}
\end{figure}

For each APOGEE spectrum, a best match within the synthetic grid is found. To
do this, interpolation within the grid, as well as an efficient search
method, are required. This is accomplished using the FORTRAN95 code 
FERRE \citep{AllendePrieto2006}\footnotemark{1},
which yields a best set of stellar
parameters. The computational resources to perform the full
7D search for $>$ 100,000 stars are substantial ($>$ 50,000 CPU
hours).  Because of this, we use a calibration subsample (see \S
\ref{sect:calibration}) to derive a relation for the microturbulent
velocity as a function of surface gravity. Figure \ref{fig:vmicro}
shows the derived microtubulent velocity as a function of surface
gravity for the calibration subsample, color-coded by metallicity.
The stars of higher metallicity that make up the bulk of the
APOGEE sample (but not the calibration sample) define a fairly
tight sequence.  We a fit a linear relation
to the subset of stars from these data with \logg $ <3.8$ 
and \mh$ >-1$ and derived the fit that was adopted for DR12: 
\begin{equation} 
v_{micro}=2.478 -0.325 \log g,
\end{equation} 
which is shown as a solid line in Figure \ref{fig:vmicro}.
We then use this relation to construct
a 6D grid, interpolating within the 7D grid to the microturbulent
velocity appropriate for the surface gravity of each grid point.
The utility of this is that it reduces the search to a 6D one,
rather than 7D, with a corresponding reduction in the time required
(by a factor of three or more) to analyze the full sample.

\footnotetext[1]{http://www.sdss3.org/svn/repo/apogee/aspcap/ferre, http://hebe.as.utexas.edu/ferre}

The full DR12 sample is then analyzed using the 6D grid to provide
the initial set of parameters: \teff, \logg, \mh, \cm,
\nm, and \am.

Given the stellar parameters, the spectra are then analyzed for
abundances of each of 15 different elements --- C, N, O, Na, Mg, Al,
Si, S, K, Ca, Ti, V, Mn, Fe, and Ni --- that have features in the
APOGEE wavelength coverage. For each element, the regions in the
spectrum that are sensitive to the abundance of that element, and
less sensitive to the abundance of other elements, are determined
from model spectra. The spectral ``windows" identified in this
way are weighted by their relative sensitivities and also by how
well a synthetic spectrum with the parameters appropriate to Arcuturus
match those of an observed Arcturus spectrum. Details of this procedure,
and the specific locations of the derived windows, 
are discussed in \citet{GarciaPerez2015}.

Given the windows for each element, each spectrum is then run
through the FERRE search code separately for each element, using
the same grid as used for the determination of the stellar
parameters. However, for the individual element abundances,
the search is done holding the stellar parameters \teff, \logg , 
$v_{micro}$ fixed, and just varying a single metallicity parameter, so this
is a one-dimensional search.  For C and N, the \cm\  and \nm\  
dimension is varied, for the $\alpha$ elements (O, Mg, Si, S, Ca, and Ti),
the \am\   dimension is varied, while for all others (Na, Al, K,
V, Mn, Fe, Ni), the \mh\  dimension is varied. Apart from C and N,
this means that more than just the single element abundance being
sought is being varied, but ASPCAP attempts to account for that
by using windows with maximal sensitivity to the element being
derived and minimal sensitivity to other abundances. This allows
us to avoid having to create a series of 15 different 7D
grids where pure individual element variations are calculated.

\subsection{FERRE output}

The full ASPCAP pipeline produces an initial set of six parameters
for each star from the initial 6D fit, plus 15 additional abundance
measurements. The raw output from the 6D fit are stored in an array
called \flag{FPARAM}, while the output from the 15 separate 1D ``window"
fits are stored in an array called \flag{FELEM}.

FERRE also produces a covariance matrix from the initial 6D fit,
which we store in a matrix called \flag{FPARAM\_COVAR}, and individual
abundance uncertainty estimates, which we store in a \flag{FELEM\_ERR}
array. Unfortunately, we have found that the formal errors returned
by FERRE seem to be significantly underestimated when they are
calculated using standard fit techniques (i.e., inverse of the
curvature matrix); this is plausibly caused
because systematic errors (e.g., in the synthetic calculations,
LSF-matching, etc.) dominate statistical ones (see 
\citealt{GarciaPerez2015} for more details). As a result, we also
determine and store a more empirical estimate of uncertainties, as
discussed below.

For each parameter and elemental abundance, we store a
bitmask flagging possible conditions encountered in the process, such
as a solution falling near the edge of the grid, and consequently deemed
to be less reliable.

The data products are discussed in more detail in \S \ref{sect:access}.
The ASPCAP software is version controlled in the SDSS-III software 
repository in the speclib, idlwrap and ferre products. 

\subsection{Versions and data releases}

There are four main stages of APOGEE data processing: visit reduction, star
combination, ASPCAP/FERRE processing, and final calibration. The software
is set up such that these four stages can be run with independent 
software versions/configurations
so that, for example, a later stage can be run without having to rerun
an earlier stage with a new software version. The output from the four
stages is saved in a series of nested subdirectories, as described in
\S \ref{sect:access}.
The names of the different versions are tied to software version numbers
of the different products.
Table \ref{tab:versions} gives the names and versions used for the APOGEE
data releases.

\begin{deluxetable*}{lllll}
\tablecaption{Software version and product names}
\tablehead{
\colhead{Data release} & \colhead{APRED\_VERS (apogeereduce)} & \colhead{APSTAR\_VERS (apogeereduce)} & \colhead{ASPCAP\_VERS (idlwrap)} & \colhead{RESULTS\_VERS (idlwrap)} \\
}
\startdata
DR10 & r3 (v2\_11) & s3 (v2\_11)   & a3 (v7)    & v304 (v15) \\
DR11 & r4 (v2\_14) & s4 (v2\_15)   & a4 (v16)    & v402 (-)   \\
DR12 & r5 (v3\_04) & stars (v3\_10) & l25\_6d (v22) & v603 (v27) \\
\enddata
\label{tab:versions}
\end{deluxetable*}

\section{Calibration and validation of DR12 parameters and abundances}
\label{sect:calibration}

Results for the stellar parameters and abundances have been calibrated
and validated based on observations of a calibration sample. This sample
includes stars in the \textit{Kepler} field that have well-characterized effective
temperatures and surface gravities, samples of stars in a
set of globular and open clusters that span a wide range of metallicity,
and a set of stars with previously-measured parameters and abundances.

\subsection{Effective temperatures}

\begin{figure}
\includegraphics[width=0.4 \textwidth]{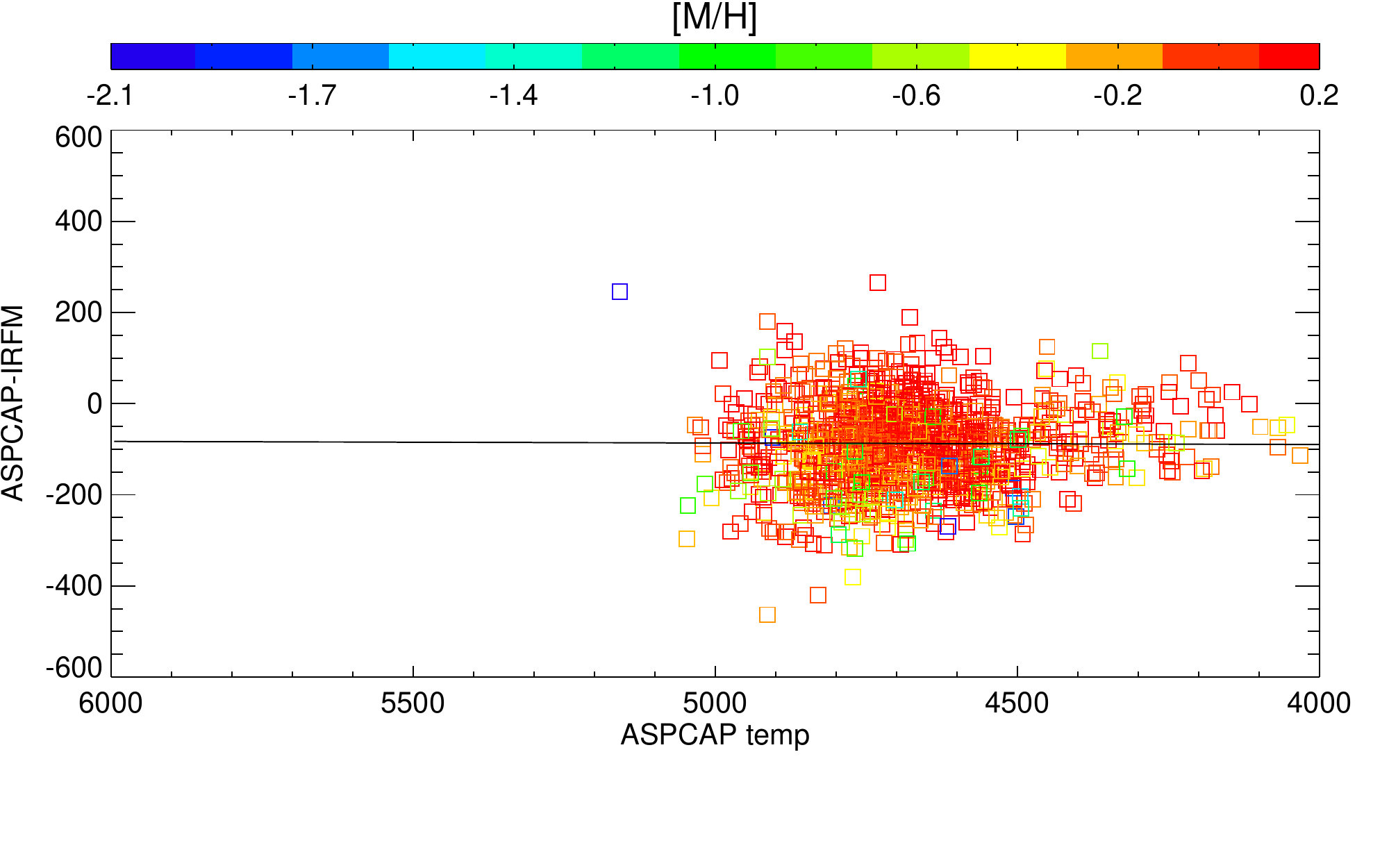}
\caption{Temperature calibration, derived from low-reddening sample with
photometric temperatures.
}
\label{fig:teff}
\end{figure}

We compare the ASPCAP spectroscopically-determined
temperatures with those obtained from the calibrated photometric relation of
\citet{GHB2009}. We chose a sample
of giants (\logg $<$ 3.8)  in the \textit{Kepler} field and adopted E(B-V) estimates as determined
from the SAGA survey \citep{Casagrande2014}. We convert $E(B-V)$ to $E(J-K)$ 
using
$$E(J-K)=(0.56+0.06 (J-K)_0)*E(B-V)$$
\citep{Bessell1998}.  We use the dereddened $J-K$ color to derive the
temperature, and adopt the raw ASPCAP metallicity for the
color-temperature relation. We restrict the comparison to stars
with ASPCAP metallicity \mh$>-1$.  Figure \ref{fig:teff} shows the
difference between the spectroscopic and photometric temperatures for
this subsample. This comparison suggests that there is an offset in
the derived effective temperature in that the raw ASPCAP temperature is
about 90 K cooler than the photometric temperatures.

We derived a calibration based on a linear fit with temperature to these data,
which yields:
\begin{equation}
T_{\rm{eff,corr}} = T_{\rm{eff,raw}} - 0.0034 (T_{\rm{eff,raw}}-4500) + 87.9
\label{eqn:teff}
\end{equation}
although the dependence of the difference on temperature is negligibly small.

The calibrated temperature provided for all giants in the DR12 release
uses the calibration from equation \ref{eqn:teff}. For giants with
effective temperatures outside the range of those used to derive the
calibration  ($3500<$\teff $<6000$), the correction at the edge of the
range was adopted, and the \flag{CALRANGE\_WARN} bit has been set in
the \flag{PARAMFLAG} for temperature (see \S \ref{sect:usingdata}).


An upper limit on the temperature uncertainty is derived from the
scatter around this relation, which gives $\sigma(T_{eff}) = 91.5$
K. This is an upper limit because there are uncertainties associated
with the photometry. On the other hand, they may be some systematic
uncertainties related with the photometric temperature relation.
This scatter around the calibration relation was adopted as the empirical
temperature uncertainty for stars in DR12.

The color-temperature relation of \citet{GHB2009} is also used to flag
objects that may have poor stellar parameters. For each object, a
temperature is determined using the color-temperature relation, the
2MASS $J-K$ color, and the estimate of the reddening used by the
targeting pipeline. This is compared with the ASPCAP-derived temperature.
When these deviate by more than 1000 K, the \flag{COLORTE\_WARN} bit
is set in the \flag{ASPCAPFLAG} (see \S \ref{sect:usingdata}); if
the temperature differ by more than 2000 K, the \flag{COLORTE\_BAD} bit
is set and the ASPCAP results are flagged as bad. Most of these are
cool dwarfs that are not fit well (in retrospect, we might have
chosen less conservative values to flag).

\subsection{Surface gravities}
\label{sect:logg}

Surface gravity is a key parameter for a number of reasons: (1) it is important for
estimating distances to survey stars, (2) it can be used to identify AGB stars,
and (3) it can have a significant effect on line strengths.

The APOGEE survey includes observations of giants that have time series
photometry with \textit{Kepler} that allows asteroseismic analysis.
In principle, such analysis very accurate surface gravities, with an
estimated uncertainty of $<<0.05$ in $\log g$ \citep{hekker2013}.

A calibration relation for surface gravity has been derived using a subset
of stars from version 7.3 of the APOKASC catalog \citep{Pinsonneault2014}.
To cover a wide range of properties, we took all stars that had inferrable
evolutionary states from \citet{Stello2013}, all stars with \logg $<$2, and
all of the small number of stars in the APOKASC catalog with [M/H] $<$-1.



\begin{figure}
\includegraphics[width=0.4 \textwidth]{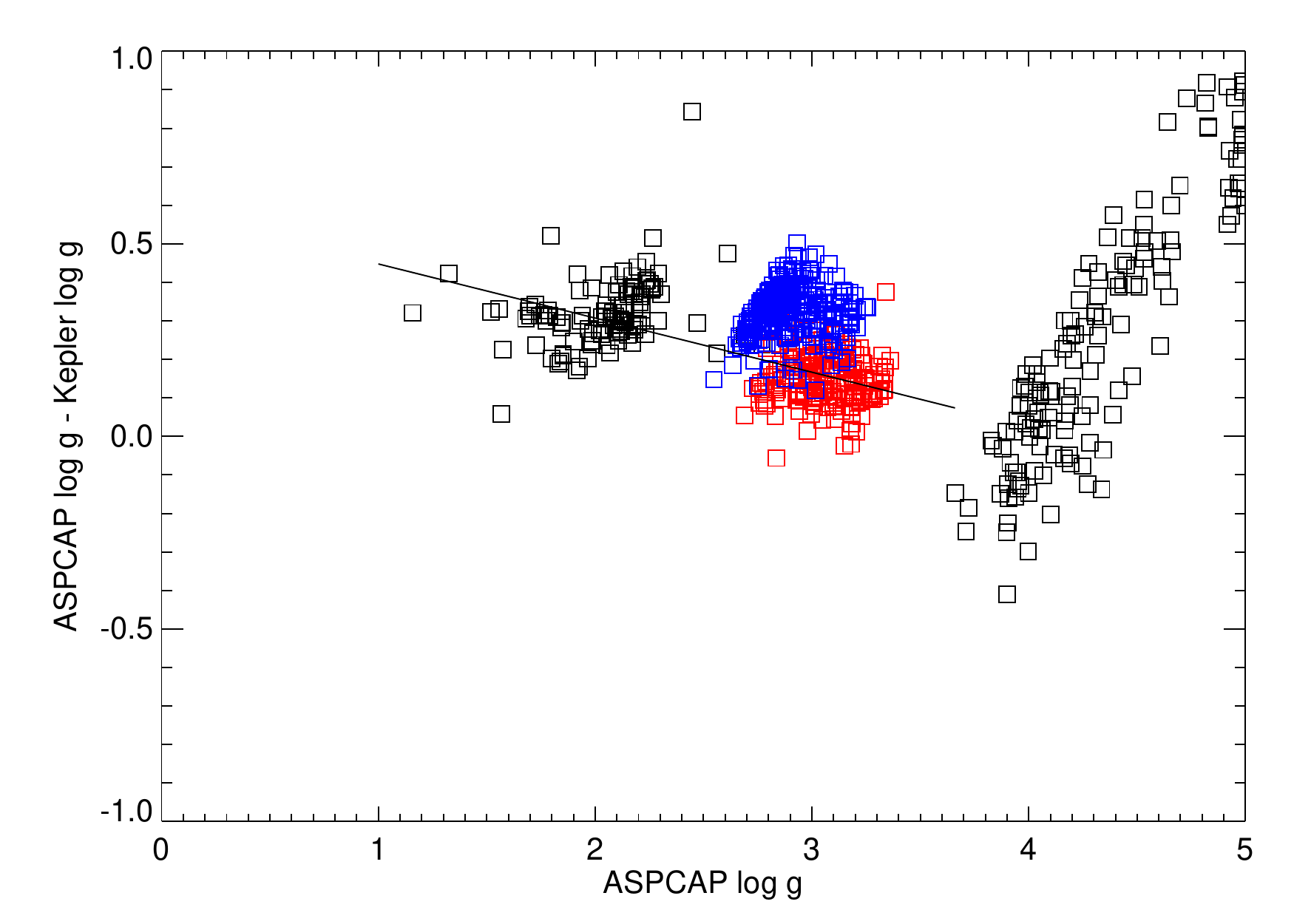}
\caption{Surface gravity calibration, derived from a sample of
stars in the \textit{Kepler} field with gravities from asteroseismic 
analysis. This analysis also provides information on the evolutionary
state: red points are hydrogen shell burning (RGB) stars, while
blue points are helium core burning (RC) stars. 
}
\label{fig:logg}
\end{figure}

Figure \ref{fig:logg} shows the difference between ASPCAP spectroscopy
$\log g$ and the asteroseismic $\log g$ from \citet{Pinsonneault2014}, as a function
of surface gravity. 
It is clear that results for stars at higher surface gravity ($\log g \geq 4$)
are poor. We expect that this is caused to a large extent by the lack
of treatment of rotation discussed above.

At lower surface gravities, there are systematic offsets between
gravities determined for RGB and RC stars; the difference between
ASPCAP and asteroseismic gravities is larger for RC stars, even for
RGB/RC pairs at the same gravity and metallicity. This has already been
noted in previous APOGEE samples \citep{Pinsonneault2014, Meszaros2013},
but the effect is still currently not well understood.
If one considers just the RGB sample, it appears that the difference
between ASPCAP and asteroseismic gravity increases to lower gravities. 
No dependence on metallicity was found.

We have chosen to adopt a surface gravity calibration using a linear
surface gravity fit to the RGB sample. If one assumes that the
asteroseismic gravities are correct, this would imply that the RC stars
in the DR12 sample will have calibrated surface gravities that are
too large by $\sim$ 0.2 in $\log g$. We have chosen this approach rather than
``splitting the difference" between the RGB and RC samples because it
is possible to distinguish the bulk of the RC stars from RGB stars
based on cuts in temperature and surface gravity (see \citealt{Bovy2014}).
Because RC stars are expected to have well-constrained luminosities, it
is possible to use these for distance estimates \citep{Bovy2014} instead
of needing to rely on the ASPCAP surface gravities. We thus choose to
adopt the calibration relation that we hope will give the best results
for RGB stars, and recommend treating the RC stars separately.
The derived calibration relation is given by:

\begin{equation}
\log g_{corr} = \log g_{raw} + 0.14 \log g_{raw} - 0.588.
\label{eqn:logg}
\end{equation}

For uncertainties in $\log g$, we adopt the observed scatter around this
calibration relation, again ignoring the RC stars. This gives an
empirical uncertainty of 0.11 in $\log g$. However, the true error
is probably more likely to be dominated by systematics, given the
\textit{Kepler} differences. We have also determined that the
surface gravity is sensitive to the details of the LSF used in
the analysis, so this may lead to systematics depending on the
exact LSF combination of each spectrum.

\subsection{Calibrated parameters and abundance determination}

Our procedure to determine elemental abundances uses the raw derived
stellar parameters when fitting synthetic spectra to the windows in
the spectra that cover features of specific elements. We choose to do this
because (1) using calibrated parameters would lead to significantly
poorer overall spectrum fits, and (2) it is possible that the raw,
``spectroscopic," parameters account to some extent for uncertainties
and approximations that are implicit in the synthetic spectra, and
so using the raw parameters provides a more self-consistent set
of abundances.

On the other hand, it is also possible that the raw parameters are
offset for other reasons, for example, if the adopted LSF leads to
incorrect surface gravities. If this were the case, then one might
argue that using the calibrated parameters for individual elemental
abundances might be more accurate, at least for features where the
strength depends strongly on surface gravity. Future work is
planned to investigate this.

\subsection{Chemical Abundances}
\label{sect:abuncal}

Our primary calibration sample for abundances is a set of globular and
open clusters spanning a wide range of metallicity, from M92 and M15
on the metal-poor end to NGC 6791 on the metal-rich end.  The set of
calibration clusters is presented in Table \ref{tab:clusters}.

We have approached calibration in two steps: an internal calibration where
we look for homogeneity of elemental abundances within clusters, and an
external calibration where we investigate the agreement of the abundances
with previously measured values.

\begin{deluxetable}{lrl}
\tablecaption{Calibration clusters}
\tablewidth{0.4 \textwidth}
\tablehead{
\colhead{cluster} & \colhead{Literature [Fe/H]} & \colhead{Reference} \\
}
\startdata
M92 &    -2.35  & \citet{Harris1996}\\
M15 &    -2.33  & \citet{Harris1996}\\
M53 &    -2.06  & \citet{Harris1996}\\
N5466 &    -2.01  & \\
N4147 &    -1.78  & \citet{Harris1996}\\
M2 &    -1.66  & \citet{Harris1996}\\
M13 &    -1.58  & \citet{Harris1996}\\
M3 &    -1.50  & \citet{Harris1996}\\
M5 &    -1.33  & \citet{Harris1996}\\
M107 &    -1.03  & \citet{Harris1996}\\
M71 &    -0.82  & \citet{Harris1996}\\
Be29 &    -0.44  & \citet{Carraro2004}\\
N2243 &    -0.35  & \citet{Jacobson2011b}\\
M35 &    -0.14  & \citet{Barrado2001}\\
N2420 &    -0.13  & \citet{Jacobson2011}\\
Pleiades &     0.03  & \citet{Soderblom2009} \\
N188 &     0.04  & \citet{Jacobson2011}\\
N2158 &     0.04  & \citet{Jacobson2009}\\
M67 &     0.06  & \citet{Jacobson2011}\\
N7789 &     0.09  & \citet{Jacobson2011}\\
N6819 &     0.16  & \citet{Bragaglia2001}\\
N6791 &     0.37  & \citet{Carraro2006}\\
\enddata
\label{tab:clusters}
\end{deluxetable}

\subsubsection{Internal calibration}
\label{sect:intcal}

For the internal calibration, we examine the ASPCAP abundances under
the assumption that clusters are chemically homogeneous for all
elements except carbon and nitrogen, which are expected to have
variations in giants because of potential mixing of CNO-processed material.
The sample was restricted to more metal-rich clusters 
because (1) multiple populations in some globular clusters
are manifest in elemental abundances (e.g., Al, Mg, Na, and O), and our
cluster sample is dominated by globular clusters at \mh $<$ -1; (2)
at lower metallicity, the lines are weaker, so trends are more
challenging to discern; and (3) the bulk of the main APOGEE sample
is at higher metallicity, so we choose to provide an internal
calibration that is most accurate for the bulk of the sample.  In
practice, we inspected the spread of each individual element within
clusters of various metallicities, and chose a lower limit for
each element to use for the internal calibration; this limit was
always between \mh of -1 and -0.6.

For the internal calibration, we look for agreement among the cluster
stars, without any constraints on what the mean abundances are.
Inspection of the raw ASPCAP abundances within each cluster suggests
that the abundances of some elements appear to have small trends with
effective temperature. To quantify this effect we fit, for each element,
for mean abundances of each cluster plus a linear temperature dependence
of the abundance:

\begin{equation}
[\rm{X_{ij}/H}] = [\rm{X_i/H}] + S_{\rm{X}} (T_{ij} - 4500 ) / 1000.
\label{eqn:intcal}
\end{equation}
where $i$ represents the cluster, and $j$ the star within the cluster,
and $S_{\rm{X}}$ the derived slope for element X.  Note that we used
\xh{X}\  for all abundances: for the elements for which ASPCAP uses
a grid dimension that is relative to the overall metal abundance
(C, N, and $\alpha$ elements), we added the ASPCAP \mh\  to get
\xh{X}. This made it easier to detect potential cluster non-members as
those stars with abundances significantly different from the cluster
averages.  We attempted to be conservative about cluster membership:
we started with stars selected to be probable cluster members (see
\citealt{Zasowski2013}), but then adopted only stars with radial
velocities within 5 km/s of our cluster mean, and then finally rejected
stars with abundance residuals more than 0.2 dex from the fit.

\begin{figure*}
\includegraphics[width=\textwidth]{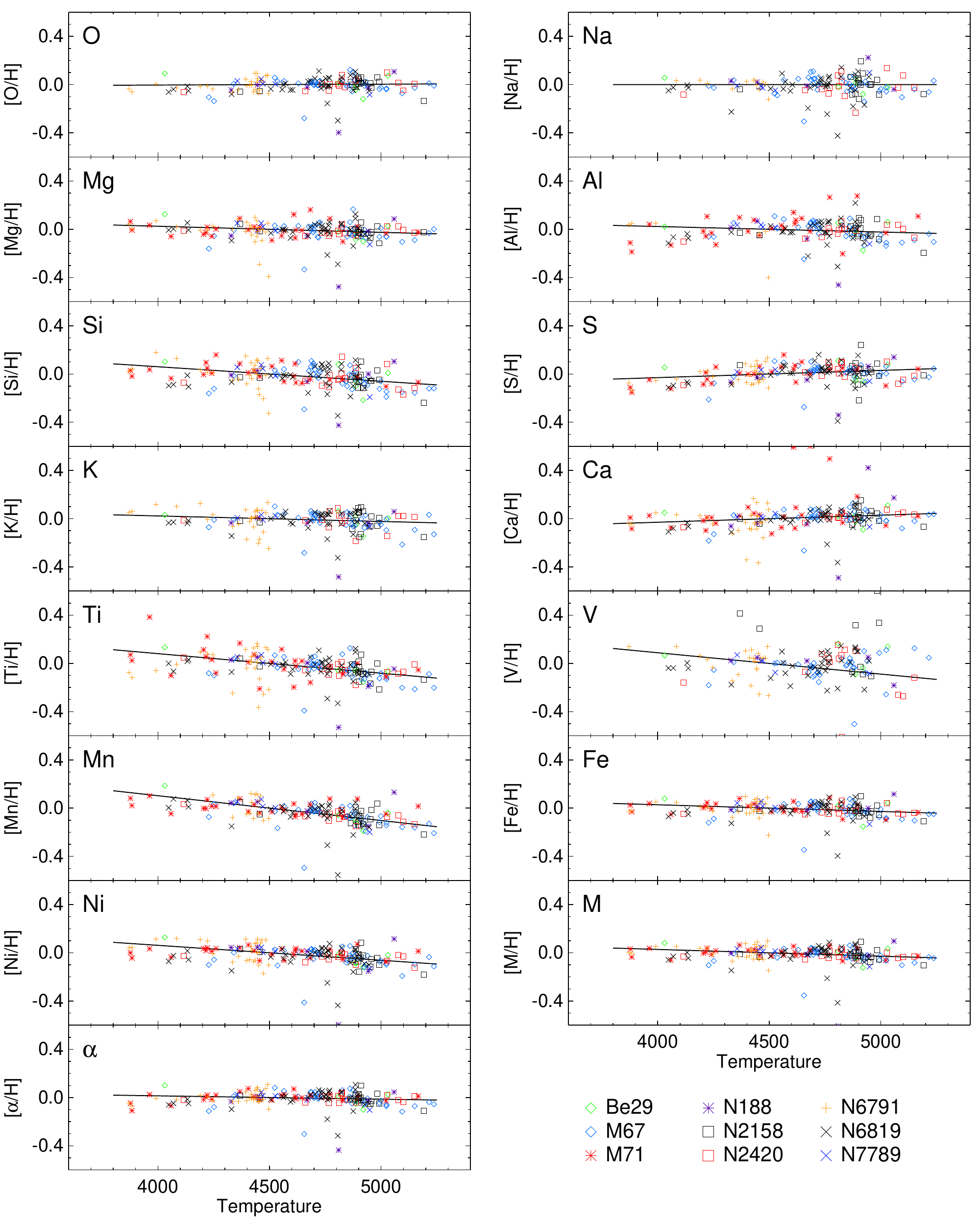}
\caption{Internal elemental abundance calibrations. Plots show deviations
of individual elements relative to the cluster means as a function of
temperature; different symbols are for different clusters, as noted
in lower right. Lines shows the adopted temperature calibrations. Carbon
and nitrogen are not shown because we cannot assume homogeneous abundances
within clusters.}
\label{fig:internal}
\end{figure*}

Results are shown in Figure \ref{fig:internal}, where abundances of
each cluster calibration star are shown relative to the derived mean
($[\rm{X_i/H}]$) of the cluster of which it is a member. Points are color-coded
by the metallicity of their respective cluster.
The small temperature trends could arise from a number
of different factors: the assumption of LTE, issues with atomic data since
different lines provide information about abundances at different
temperatures, etc. The data suggest that a linear temperature term is
adequate to describe the trends for  most elements.

For DR12, we have adopted the derived temperature trends for an
internal calibration of the abundances for all elements except carbon
and nitrogen. The slopes of these temperature terms ($S_{\rm{X}}$) are given
in Table \ref{tab:tempslopes}.  We arbitrarily define the abundance of
a given element as the abundance at \teff = 4500 K.
While is it possible that the derived abundances of carbon and nitrogen may
have temperature dependencies, we cannot discern these from the cluster
data internally.

We apply the temperature corrections as derived from the cluster sample
to the DR12 sample. Because the clusters only provide stars with 
$3800<$\teff$<5250$K, we do not extrapolate the linear relation outside this
range. Instead, we use the correction at $T_{eff}=5250$K for all stars with
$T_{eff}>5250$K, and the correction at $T_{eff}=3800$K for all stars with $T_{eff}<3800$K. 
For those giants outside the temperature range of the calibrator, we set
the \flag{CALRANGE\_WARN} bit in the abundance masks.

In addition, because the calibration is derived only from giants, and
because of other issues noted above for dwarfs, we do not apply it
to stars with $\log g > 3.8$. For these higher surface gravity
stars, we do not provide calibrated abundances at all (the values
are set to -9999) and we set the \flag{CALRANGE\_BAD} bit in the abundance
masks. The raw ASPCAP parameters are still provided.

\begin{deluxetable}{lr}
\tablecaption{Internal temperature calibration slopes}
\tablewidth{0.2 \textwidth}
\tablehead{
\colhead{Element} & \colhead{slope per 1000 K}  
}
\startdata
C  &    \nodata \\
N  &    \nodata \\
O  &   +0.007 \\
Na &   -0.001 \\
Mg &   -0.051 \\
Al &   -0.046 \\
Si &   -0.121 \\
S  &   +0.060 \\
K  &   -0.046 \\
Ca &   +0.059 \\
Ti &   -0.162 \\
V  &   -0.177 \\
Mn &   -0.206 \\
Fe &   -0.055 \\
Ni &   -0.124 \\
\mh &   -0.056 \\
\am &   -0.028 \\
\enddata
\label{tab:tempslopes}
\end{deluxetable}

\subsubsection{Empirical abundance precision}
\label{sect:errors}

The individual abundances of stars in clusters provide an excellent
data set to assess the precision of our derived abundances, by
using the scatter within clusters as an estimate of the abundance
uncertainties. This scatter is a function of both temperature and
metallicity, which is as expected because lines are weaker at lower metallicity,
and change strength with temperature.  In addition, we see that scatter
decreases with increasing signal-to-noise.

To derive empirical uncertainties, we calculated the spread of the
abundance within each cluster in bins of width 250K in temperature
and 50 in S/N (putting all observations with S/N$>200$ into the highest
S/N bin).  To achieve a greater range in S/N, we included abundance measurements
for cluster stars derived from individual visits in addition to those
from the combined spectra.  We combined results for all clusters in bins
of 0.5 dex in metallicity, and then fit a simple functional form to the
uncertainty in each element:

\begin{multline}
\ln(\sigma_X) = A_X + B_X (T_{eff} - 4500) + \\ C_X [M/H] + D_X (S/N - 100)
\label{eqn:error}
\end{multline}
where we fit for the natural log to ensure that the uncertainty never goes
negative (i.e., outside of the range of the calibration cluster data).

\begin{deluxetable*}{lrrrrcc}
\tablecaption{Empirical uncertainty relation coefficients and values}
\tablewidth{0pt}
\tablehead{
\colhead{Element} & \colhead{A}  & \colhead{B} & \colhead{C} & \colhead{D} & \colhead{$\sigma$(\teff$=4500$,\mh$=0$,S/N$=100)$} & \colhead{``Global'' uncertainty}
}
\startdata
C  &   -3.350 &    0.769 &   -0.919 &   -0.066 &  0.035 & \nodata\\
N  &   -2.704 &    0.291 &   -0.591 &   -0.078 &  0.067 & \nodata\\
O  &   -3.649 &    0.670 &   -0.614 &   -0.093 &  0.026 & 0.050\\
Na &   -2.352 &   -0.002 &   -0.915 &   -0.263 &  0.095 & 0.064 \\
Mg &   -3.537 &    0.263 &   -0.825 &   -0.297 &  0.029 & 0.053 \\
Al &   -2.764 &    0.471 &   -0.868 &   -0.162 &  0.063 & 0.067 \\
Si &   -3.150 &    0.383 &   -0.224 &   -0.105 &  0.043 & 0.077 \\
S  &   -3.037 &    0.507 &   -0.625 &   -0.299 &  0.048 & 0.063 \\
K  &   -2.770 &    0.216 &   -0.667 &   -0.275 &  0.063 & 0.065 \\
Ca &   -3.226 &    0.284 &   -0.879 &   -0.429 &  0.040 & 0.059 \\
Ti &   -3.186 &    0.657 &   -0.819 &   -0.068 &  0.041 & 0.072 \\
V  &   -1.608 &    0.900 &   -0.400 &   -0.418 &  0.200 & 0.088 \\
Mn &   -3.031 &    0.639 &   -0.661 &   -0.326 &  0.048 & 0.061 \\
Fe &   -3.357 &    0.098 &   -0.303 &   -0.071 &  0.035 & 0.053 \\
Ni &   -3.153 &    0.135 &   -0.493 &   -0.185 &  0.043 & 0.060 \\
\mh &  -3.603 &    0.109 &   -0.433 &    0.039 &  0.027 & 0.049 \\
\am &  -4.360 &    0.060 &   -0.848 &   -0.096 &  0.013 & 0.048 \\
\enddata
\label{tab:errpar}
\end{deluxetable*}

Table \ref{tab:errpar} gives the coefficients for these empirical error
estimates for all of the abundances. The second-to-last column gives
the empirical error estimate at \teff $ = 4500$K, solar metallicity, and
$S/N=100$. For all elements except vanadium, the observed scatter is less
than 0.1 dex, and for many elements, it is below 0.05 dex. Generally,
the errors increase with increasing temperature, decreasing metallicity,
and decreasing S/N.

These empirical errors, which are distributed in the APOGEE data release
(along with the raw FERRE errors), capture the scatter in elemental
abundances observed in clusters over a narrow range of effective
temperature. Over a broader range of temperature, they do not capture the
degree to which our simple, metallicity-independent, linear temperature
calibration relation fits the data. Since the observed stars in any given
cluster do not usually span a large temperature range, we estimate the
overall uncertainty in abundances, across a broad range of temperature,
by measuring the scatter around the calibration relations of Figure
\ref{fig:internal}.  This is presented in the final column of Table
\ref{tab:errpar}.  These ``global" uncertainties do not capture the
likely dependence of the precision on temperature, metallicity, and
signal-to-noise. They represent a more conservative estimate of the
internal uncertainties.

The scatter within clusters indicates that the precision of our
abundances is less than 0.1 dex for all elements, with the possible
exception of vanadium. For some abundances, especially if considering
the sample over a restricted temperature range and high S/N, the
precision can be better than 0.05 dex.

\subsubsection{External calibration}

\label{sect:extcal}

\begin{figure}[t]
\includegraphics[width=0.5 \textwidth]{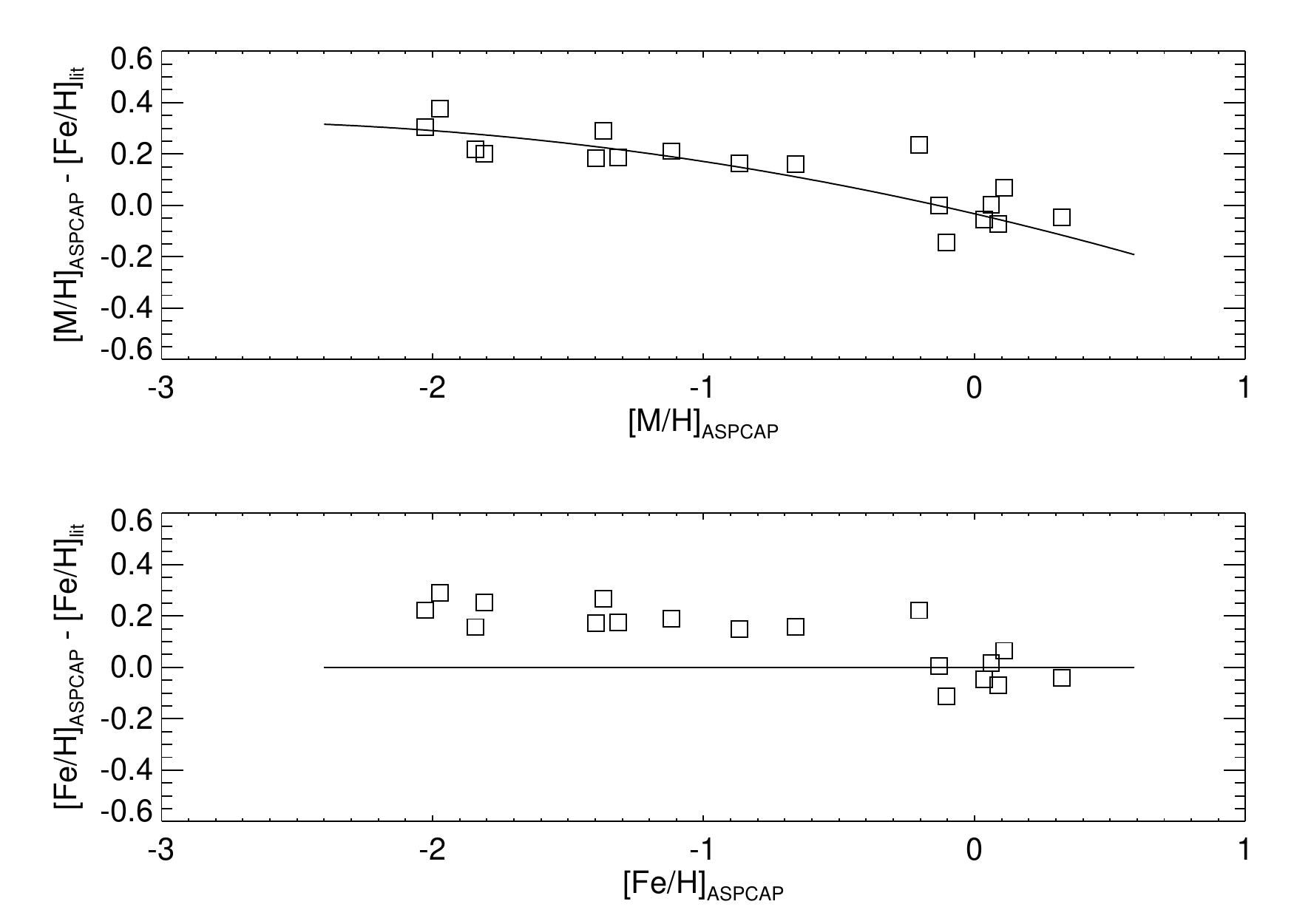}
\caption{External elemental abundance calibration.}
\label{fig:extcal}
\end{figure}

Table \ref{tab:clusters} gives the adopted literature values for \feh, 
along with references, for the calibration cluster sample. The difference
between the derived ASPCAP values and the literature values for both the 
parameter \mh\  and the abundance \feh\  are shown in Figure 
\ref{fig:extcal}. We adopt the literature \feh\ as a reference for
our \mh\ values because our method allows for variations in $\alpha$ elements,
carbon, and nitrogen separately; implicit in this comparison is 
the assumption that the Fe abundance drives the overall metallicity in 
the APOGEE spectra.

The derived ASPCAP $[M/H]$ and $[Fe/H]$ are similar to the literature
values near solar metallicity, but are $\sim0.2-0.3$ dex more metal-rich
than the literature for very metal-poor clusters. This is similar to
what was seen for the DR10 data release \citep{Meszaros2013}. We adopted
a quadratic relation in metallicity to derive calibration relations:

\begin{multline}
[M/H]_{corr} = [M/H]_{raw} + 0.0260 + \\ 0.255 [M/H]_{raw} + 0.062 [M/H]_{raw}^2 
\label{eqn:extcal}
\end{multline}
\begin{multline}
[Fe/H]_{corr} = [Fe/H]_{raw} + 0.0326 + \\ 0.245 [M/H]_{raw} - 0.042 [M/H]_{raw}^2
\label{eqn:extcal2}
\end{multline}

We have chosen to apply the calibration above to the \mh\  parameter,
but not to the \feh\  abundances, given the impossibility of calibrating
all of the abundances in a homogeneous fashion. Most
users will prefer to work with \xfe\  by subtracting the \feh\  from
\xh{X}, and in many cases systematic errors in \feh\  and \xh{X}\  will partly
cancel out.  If we were to apply the external calibration to \feh,
it would have the effect of increasing the \xfe\  ratios of all
elements towards lower metallicity.  Since we do not know the source
of the \feh\  discrepancy at low metallicity, it is difficult to
judge whether it is likely to affect all elements similarly, or
whether it is specific to iron. As a result, we choose to leave all
of the \xfe\  ratios on the native, consistent, ASPCAP scale.


%

The main issue addressed by the calibration relation is at the metal-poor
end. If one were to ignore all clusters with \feh$<-0.5$, the remaining
metal-rich clusters show little evidence for an offset or a trend
with metallicity. However, the simple quadratic form of the relation
adopted to address the issues at \feh$<-0.5$ results in a relation that
changes by almost 0.3 dex at $-0.5<$\feh$<0.5$.  As a result, there are
non-negligible differences between the calibrated and uncalibrated \mh\
in this regime, and the uncalibrated value may be just as good as the
calibrated ones for clusters with \mh$>-0.5$. 

\subsection{The abundance scale}
\label{sect:refcal}

The ASPCAP abundances are presented as relative to hydrogen in the DR12 database in the usual bracket notation:

\begin{equation}
[X/H] = \log (n_X/n_H) - \log(n_X/n_H)_\sun
\end{equation}
where $n_X$ and $n_H$ are the number density of
nuclei of element $X$ and hydrogen, respectively. But some caveats apply. The
abundances are not truly differential to the values that APOGEE
derives for the Sun. Solar abundances
are adopted from \citet{Asplund2005} and used for computing
model atmospheres \citep{Meszaros2012} and fine-tuning the line list
\citep{Shetrone2015}. The line list used for spectral
synthesis includes, in addition to laboratory and theoretical transition
probabilities and damping constants, modifications to provide a compromise
match to both the solar spectrum and Arcturus. However, only atomic
lines were adjusted, and even for those, only within limits. 

Therefore, the ASPCAP abundances are not strictly differential to those of
the Sun.  Given the way the line lists were determined, the
existence of the temperature trends in the raw abundances, and the lack
of external calibration (apart from that for \mh), APOGEE defines
its own relative abundance scale.  
We consider the accuracy of this scale by several external comparisons,
including observations of the solar spectrum and of Arcturus, observations
of stars with independently-measured abundances, and consideration of the
locus in paramters and abundances of the entire APOGEE sample.


\subsubsection{Solar and Arcturus abundances}

Observations of the solar spectrum, obtained by observing the
asteroid Vesta, and of Arcturus were made using the NMSU 1m
feed to the APOGEE spectrograph. These spectra are of particular
interest because their abundances set the standard for many abundance
studies, they have high resolution spectral atlases, and because
the combination of the abundances and high resolution spectra 
were used to tune the APOGEE line list.



However, the internal calibration that depends on the ASPCAP temperature
was derived from the cluster sample which is significantly cooler
than the Sun.  Additionally, the calibration was derived from giants,
not dwarfs, so the applicability of the calibration to the solar
abundances is debatable, and this complicates the solar comparison.
Table \ref{tab:solar} presents the solar abundances derived from
the Vesta spectrum, showing both the uncalibrated abundances and the
calibrated abundances determined by applying the internal calibration
correction appropriate for 5250 K, which is the high temperature end of
the calibration sample.

In general, the observed uncalibrated solar abundances are within the
expected errors, except for magnesium, sulfur, and vanadium.  Applying
the calibration (again, outside of its derived range) slightly improves
some of the abundances, but makes the magnesium, silicon, manganese, and
vanadium abundances worse. Systematic offsets from the solar abundance
all appear to be within 0.1 apart from sulfur and vanadium, and perhaps
silicon (if one considers the calibrated value).

\begin{deluxetable}{lrr}
\tablecaption{Solar parameters and abundances from Vesta}
\tablehead{
\colhead{Element}&\colhead{Uncalibrated ASPCAP}&\colhead{Calibrated ASPCAP}
}
\startdata
\teff  &   5728.$\pm$    91.5&   5812.$\pm$    91.5\\
\logg  &    4.54$\pm$    0.11&    4.49$\pm$    0.11\\
\xh{M} &    0.03$\pm$    0.03&    0.11$\pm$    0.03\\
\am    &    0.00$\pm$    0.01&    0.02$\pm$    0.01\\
\xh{C} &    0.06$\pm$    0.07&    0.06$\pm$    0.07\\
\xh{N} &    0.10$\pm$    0.06&    0.10$\pm$    0.06\\
\xh{O} &    0.03$\pm$    0.01&    0.02$\pm$    0.01\\
\xh{Na}&    0.03$\pm$    0.03&    0.03$\pm$    0.03\\
\xh{Mg}&    0.07$\pm$    0.03&    0.11$\pm$    0.03\\
\xh{Al}&   -0.03$\pm$    0.01&    0.00$\pm$    0.01\\
\xh{Si}&    0.03$\pm$    0.03&    0.12$\pm$    0.03\\
\xh{S} &    0.26$\pm$    0.04&    0.22$\pm$    0.04\\
\xh{K} &   -0.07$\pm$    0.03&   -0.04$\pm$    0.03\\
\xh{Ca}&    0.02$\pm$    0.07&   -0.03$\pm$    0.07\\
\xh{Ti}&   -0.07$\pm$    0.04&    0.05$\pm$    0.04\\
\xh{V} &    0.11$\pm$    0.05&    0.24$\pm$    0.05\\
\xh{Mn}&    0.04$\pm$    0.03&    0.19$\pm$    0.03\\
\xh{Fe}&    0.01$\pm$    0.07&    0.05$\pm$    0.07\\
\xh{Ni}&    0.01$\pm$    0.15&    0.10$\pm$    0.15\\
\enddata
\label{tab:solar}
\end{deluxetable}

Table \ref{tab:arcturus} presents the ASPCAP Arcturus abundances, along
with several different comparison values: the abundances that were
adopted to tune the line list, a set of abundances derived from a
high-resolution near-IR FTS spectrum by \citet{Smith2013} using a
similar (but not identical) linelist to the current APOGEE linelist,
and a set of abundances derived from atomic transitions in the optical
spectrum by \citet{Ramirez2011}.

\begin{deluxetable*}{lrrrrrrrr}[t]
\tablecaption{Arcturus parameters and abundances }
\tablehead{
\colhead{Element}&\colhead{ASPCAP}&\colhead{Adopted}&\colhead{Smith}&\colhead{Ramirez}&\colhead{ASPCAP [X/Fe]}&\colhead{Adopted [X/Fe]}&\colhead{Smith [X/Fe]}&\colhead{Ramirez [X/Fe]}
}
\startdata
\teff& 4295.93$\pm$   91.47& & 4275.00 & 4286.00&&&&\\
\logg&    1.71$\pm$    0.11&  &   2.10 &    1.66&&&&\\
\mh&   -0.64$\pm$    0.03& &   -0.40&       ...&&&&\\
\am&    0.20$\pm$    0.02&  ...&    0.40&&&&&\\
\xh{C }&   -0.43$\pm$    0.04&   -0.43&   -0.43&   -0.09&    0.15&    0.09&    0.04&    0.43\\
\xh{N }&   -0.52$\pm$    0.08&   -0.14&   -0.14&     ...&    0.05&    0.38&    0.33&     ...\\
\xh{O }&   -0.32$\pm$    0.03&   -0.02&   -0.02&   -0.02&    0.26&    0.50&    0.45&    0.50\\
\xh{Na}&   -0.61$\pm$    0.12&   -0.52&     ...&   -0.41&   -0.04&    0.00&     ...&    0.11\\
\xh{Mg}&   -0.40$\pm$    0.03&   -0.12&   -0.38&   -0.15&    0.18&    0.40&    0.09&    0.37\\
\xh{Al}&   -0.32$\pm$    0.07&   -0.22&   -0.21&   -0.18&    0.26&    0.30&    0.26&    0.34\\
\xh{Si}&   -0.33$\pm$    0.04&   -0.12&   -0.39&   -0.19&    0.25&    0.40&    0.08&    0.33\\
\xh{S }&   -0.41$\pm$    0.04&   -0.12&     ...&     ...&    0.17&    0.40&     ...&     ...\\
\xh{K }&   -0.54$\pm$    0.06&   -0.52&   -0.29&   -0.32&    0.04&    0.00&    0.18&    0.20\\
\xh{Ca}&   -0.51$\pm$    0.04&   -0.42&   -0.47&   -0.41&    0.07&    0.10&    0.00&    0.11\\
\xh{Ti}&   -0.52$\pm$    0.05&   -0.12&   -0.31&   -0.25&    0.06&    0.40&    0.16&    0.27\\
\xh{V }&   -0.77$\pm$    0.13&   -0.52&   -0.39&   -0.32&   -0.20&    0.00&    0.08&    0.20\\
\xh{Mn}&   -0.59$\pm$    0.04&   -0.52&   -0.53&   -0.73&   -0.02&    0.00&   -0.06&   -0.21\\
\xh{Fe}&   -0.58$\pm$    0.04&   -0.52&   -0.47&   -0.52&    0.00&    0.00&    0.00&    0.00\\
\xh{Ni}&   -0.51$\pm$    0.04&   -0.52&   -0.46&   -0.46&    0.07&    0.00&    0.01&    0.06\\
\enddata
\label{tab:arcturus}
\end{deluxetable*}


Assessing the quality of the comparison is challenging because there is not a 
consensus in the literature for all Arcturus abundances.  Note that we do
not necessarily expect the ASPCAP results to match the abundances that
were adopted to tune the line list, because the line list tuning involved
making compromises between the solar spectrum and the Arcturus spectrum
(with the latter receiving lower weight, \citealt{Shetrone2015}) and
only allowed atomic transition probabilities to vary within limits.

Several different potential issues can be noted.  For carbon, the near-IR
abundances agree, but differ significantly from the \citet{Ramirez2011}
value; however, \citet{Sneden2014} find a value much closer to the
near-IR values.  For oxygen, the ASPCAP value differs significantly
from either the alternate near-IR or the optical abundance, and seems
quite low given that Arcturus has generally been found to be enhanced
in $\alpha$-element abundances. This may partly be understood because
oxygen is primarily derived from OH lines, which are quite sensitive
to temperature (see \citealt{Smith2013}) and the raw Arcuturus ASPCAP
temperature (4207K) is lower than the literature values (here is a case
where one might have gotten better results if the calibrated temperatures
were used for the abundances).  For other $\alpha$ elements (Mg, Si,
Ca), the ASPCAP values agree fairly well with the \citet{Smith2013}
values, but are lower than the \citet{Ramirez2011} values, suggesting
less overall $\alpha$ enhancement.

\subsubsection{Other individual star comparisons}

In addition to Arcturus and Vesta, we obtained NMSU 1m + APOGEE
observations of several other bright stars ($\delta$ Oph, $\beta$ And,
$\mu$ Leo) for which \citet{Smith2013} have derived abundances from
high-resolution near-IR spectra obtained with an FTS spectrograph and
using the ASPCAP line list.  We also have observations of a number
of nearby stars with published abundances from optical analyses by
\citet{Reddy2006}, \citet{Reddy2003}, and \citet{Bensby2014}; because 
these stars are
bright, observations were also all obtained with the NMSU 1m + APOGEE.
However, these are mostly warmer stars with higher surface gravity
and fall outside the range of most APOGEE main sample stars and thus
outside the range where we have internally calibrated the abundances.
We have also compiled a set of abundances of globular cluster stars from
the literature (see \citealt{Meszaros2015} for more details), to which we
can compare the APOGEE/ASPCAP abundances.

\citet{Meszaros2015} have derived abundances from APOGEE spectra for a
number of these globular cluster members. In addition, \citet{Cunha2015}
have derived Na, O and Fe abundances from manual analysis (i.e.,
doing synthesis around individual lines) of APOGEE spectra for stars
in the cluster NGC 6791, so these provide an opportunity to test the
APOGEE/ASPCAP abundances against those derived by more traditional
methods from the same spectra.  However, these independent analyses of
the APOGEE spectra were done using a slightly different version of the
APOGEE linelist, so exact agreement is not expected (but the changes
should be small, apart from K).

\begin{figure}
\includegraphics[width=0.5 \textwidth]{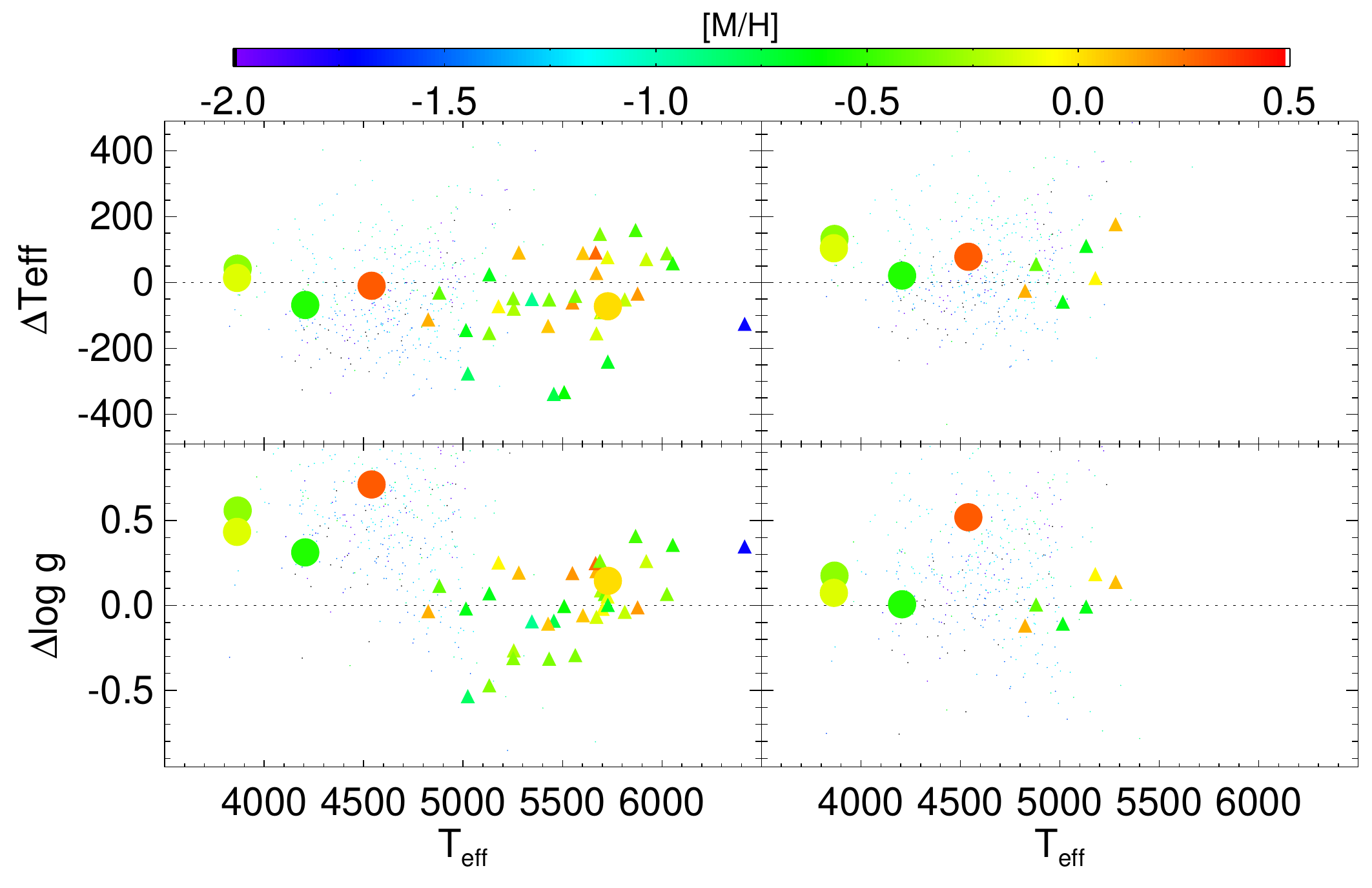}
\caption{Comparison with literature spectroscopic effective 
temperatures and surface gravities. Left panel is uncalibrated ASPCAP results; right is for calibrated results. Differences are ASPCAP minus reference values. Large circles are abundances
from FTS spectra \citep{Smith2013}, filled triangles from optical spectra and analysis, open triangles
from independent analysis of APOGEE spectra \citep{Cunha2015}.}
\label{fig:ref1}
\end{figure}

\begin{figure}
\includegraphics[width=0.5 \textwidth]{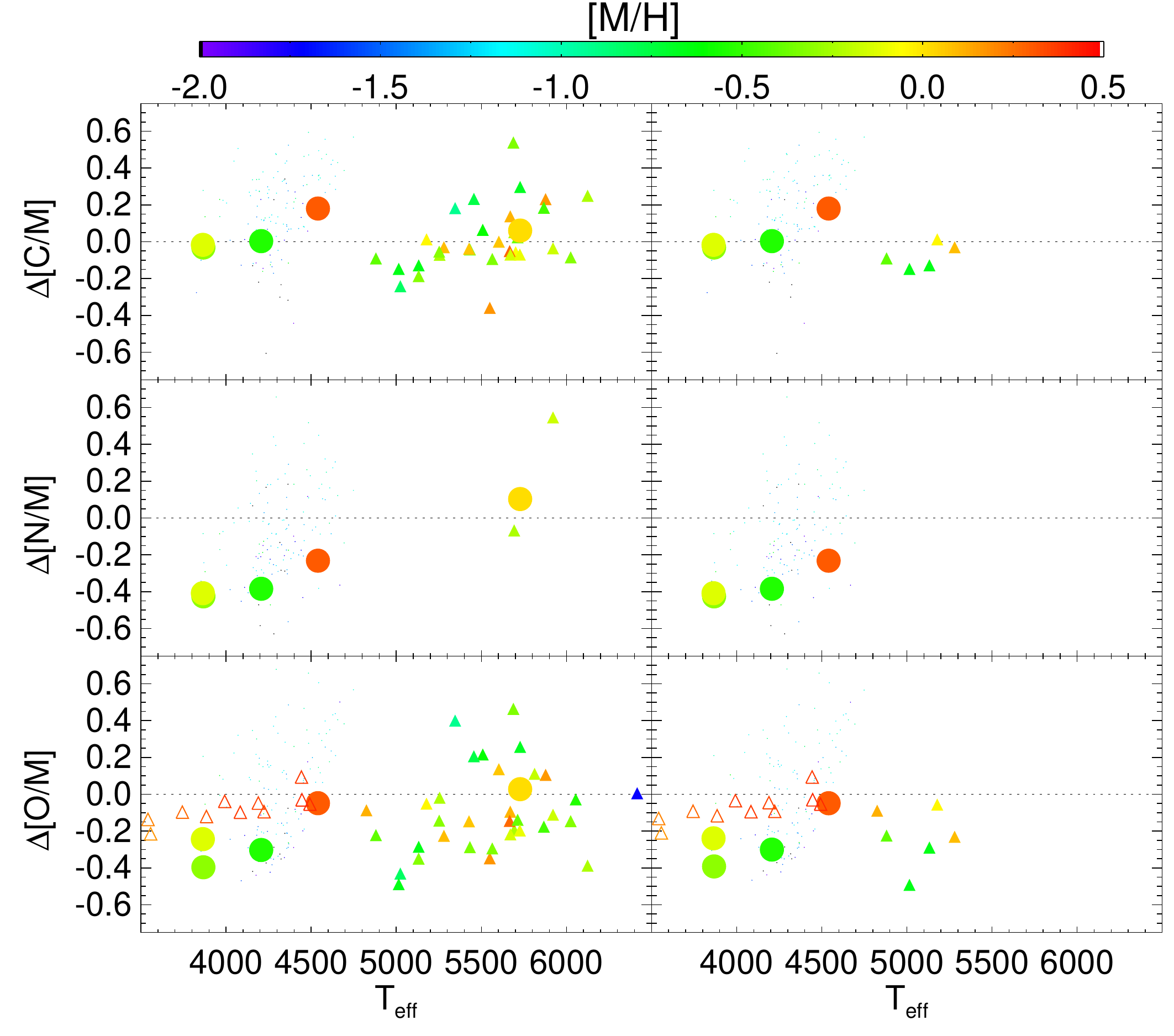}
\caption{Same as Figure \ref{fig:ref1} for CNO abundances.}
\label{fig:ref2}
\end{figure}

\begin{figure}
\includegraphics[width=0.5 \textwidth]{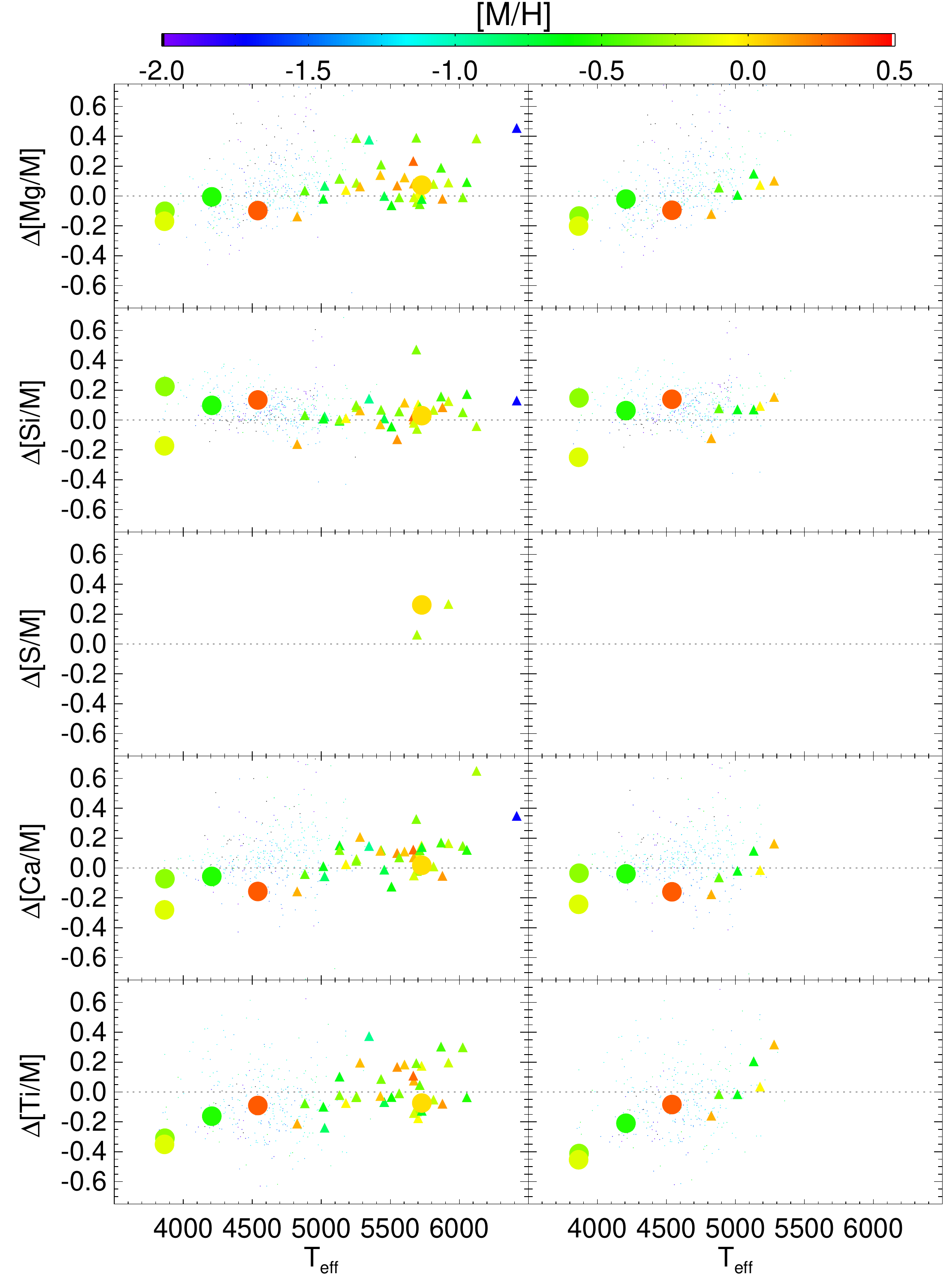}
\caption{Same as Figure \ref{fig:ref1} for $\alpha$-element abundances (Mg, Si, S, Ca, Ti. }
\label{fig:ref3}
\end{figure}

\begin{figure}
\includegraphics[width=0.5 \textwidth]{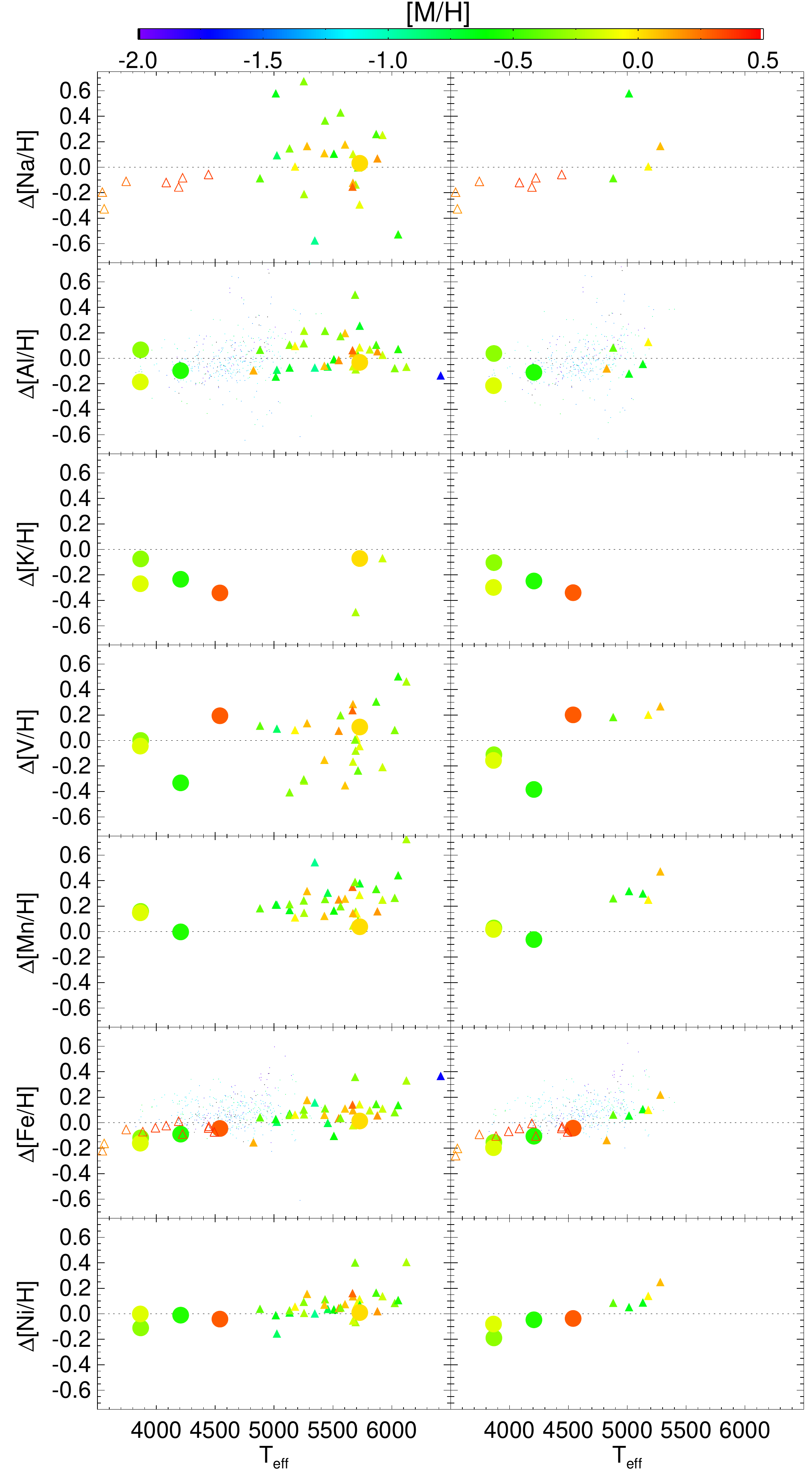}
\caption{Same as Figure \ref{fig:ref1} for Na, Al, K, V, Mn, Fe, and Ni.}
\label{fig:ref4}
\end{figure}

Results for all of these samples are shown in Figure \ref{fig:ref1},
\ref{fig:ref2}, \ref{fig:ref3}, and \ref{fig:ref4} for stellar
parameters (\teff\  and \logg), CNO, $\alpha$-element, and
other elements, respectively. In these plots, uncalibrated
parameters/abundances are shown in the left panels, and calibrated
parameters/abundances are shown on the right, both plotted as a function
of effective temperature. Since, as discussed
above, stars with higher surface gravity are not calibrated, most
of the warmer stars are missing for the calibrated parameters/abundances.
In all plots, the points are color-coded by the derived overall metallicity,
with blue representing metal-poor stars and red metal-rich ones.
Different symbols represent different comparison samples:
solid circles are the results from the FTS
spectra, solid triangles from the optical analyses of solar neighborhood
stars. Open triangles show the independent analysis for NGC 6791,
and the small dots show the results for the independent analysis of
globular cluster stars.  Although it is challenging to interpret all of
the comparisons because the comparison samples are heterogeneous and
each have their own associated uncertainties, a few potential issues
can be noticed.

%

From Figure \ref{fig:ref1}, the ASPCAP carbon abundances for giants seem
to be in reasonable agreement with literature values. ASPCAP nitrogen
abundances seem to be significantly lower, and oxygen abundances
moderately lower.

The $\alpha$ element abundances in Figure \ref{fig:ref2} suggest that Mg
compares fairly well, that Si is a bit high in the ASPCAP abundances,
and that Ca and Ti are a bit low. Titanium results may indicate a trend
with temperature; as discussed below, there is other indication that
ASPCAP titanium abundances may have significant issues that are not
currently fully understood.

For the metal-poor globular cluster stars, the ASPCAP temperatures,
surface gravities, and abundances appear offset from the literature
values and, in fact, even from the values measured from the independent
analyses of the APOGEE spectra. We suspect that this is related to
non-standard abundance patterns in some of the globular cluster
stars, e.g., oxygen abundances that do not scale with other $\alpha$
elements. The assumption that $\alpha$ elements vary in lockstep
made during the initial stellar parameter determination may lead to
stellar parameters that are systematically offset, which is evident
in the comparison of effective temperatures for the globular cluster
stars. The systematic offsets with the stellar parameters may then lead
to systematic offsets in some of the individual abundances. In addition,
since features such as CN and CO can be very weak in metal-poor stars,
the adopted scheme of allowing \cm\  and \nm\  to vary with the stellar
parameters may cause problems.

Overall, these comparisons suggest that there may be non-negligble
systematic offsets betweeen ASPCAP abundances and literature abundances
for some elements. Such offsets are perhaps not uncommon in abundance
analysis where derived values have not been adjusted relative to those
obtained from a source of similar stellar parameters (e.g., the Sun for
analysis of warmer dwarfs). Nonetheless, they do indicate that, in
an external sense, the accuracy of ASPCAP abundances may be 
poorer (0.1-0.2 dex) than the internal precision.


%


\subsection{Sample characteristics}

In this section, we provide information about the global distribution of
properties across the entire APOGEE sample, to help to indicate the quality
of derived parameters and abundances across the large parameter space of the
APOGEE stars.

\subsubsection{Quality of the fits}

\begin{figure}
\includegraphics[width=0.5 \textwidth]{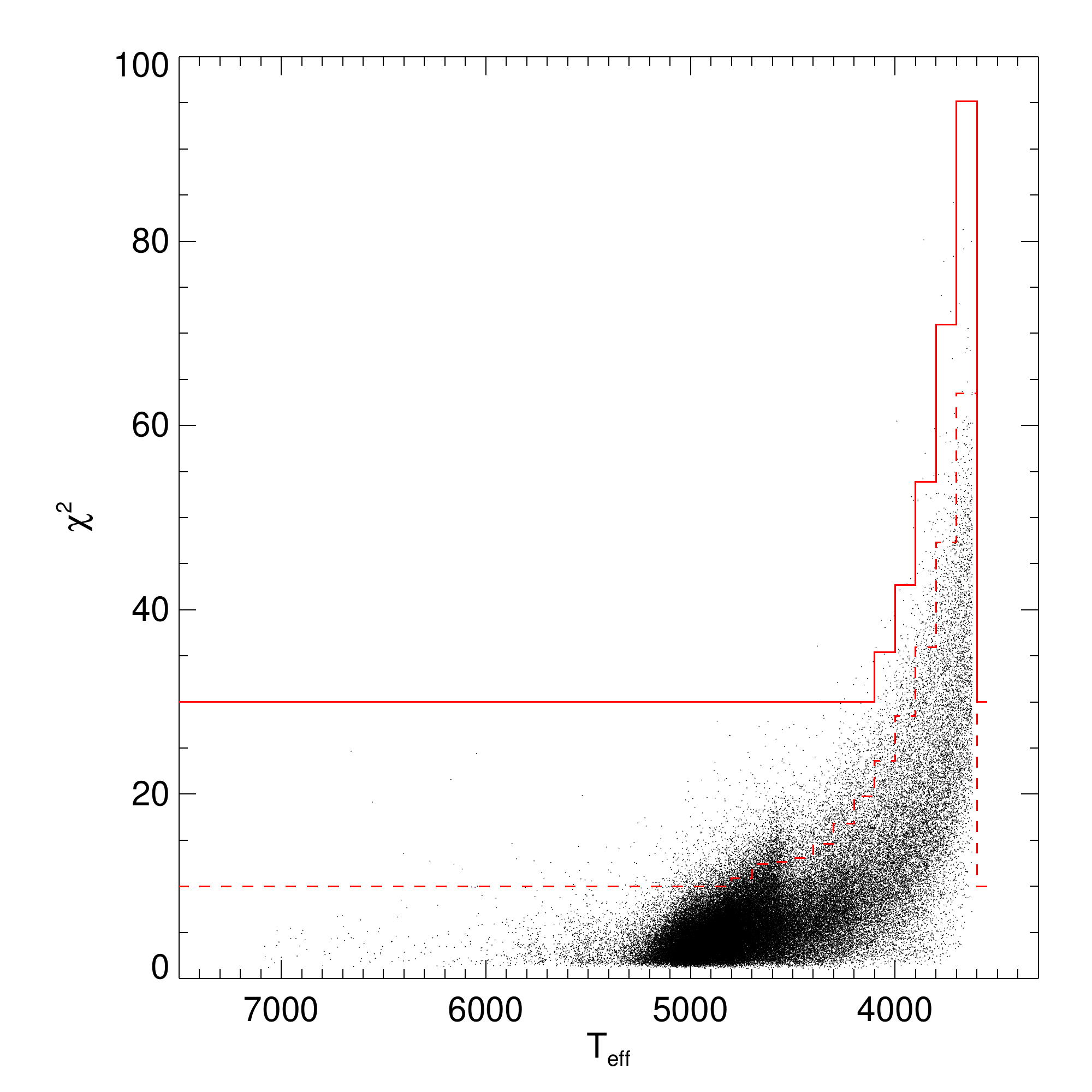}
\caption{$\chi^2$ distribution for full APOGEE giant sample. Lines
show the limits above which the \flag{CHI2\_WARN} (dashed) and 
\flag{CHI2\_BAD} (solid) bits are set.}
\label{fig:chi2}
\end{figure}

For each stellar spectrum that is fit by ASPCAP, FERRE returns a $\chi^2$ of the
fit. The $\chi^2$ distribution for the entire calibrated APOGEE DR12
sample is plotted in Figure \ref{fig:chi2}. It is clear that 
the quality of the fits are poorer for cooler stars. This is perhaps
not unexpected as the number and depth of spectral features generally
increase at cooler temperatures, so any problems, e.g., with the line list,
will be amplified at cooler temperatures.

These results suggest that there is probably more uncertainty in the
ASPCAP results at cooler temperatures.
For each star, if the $\chi^2$ value is larger than the typical value
at the temperature of the star, a bit is set in the \flag{ASPCAPFLAG} 
(see \S \ref{sect:aspcapbitmasks}):
the dashed and solid lines in Figure \ref{fig:chi2} show the levels
above which the \flag{CHI2\_WARN} and \flag{CHI2\_BAD} bits are set,
respectively.

\subsubsection{Stellar parameters}

\begin{figure*}[h]
\includegraphics[width=0.9 \textwidth]{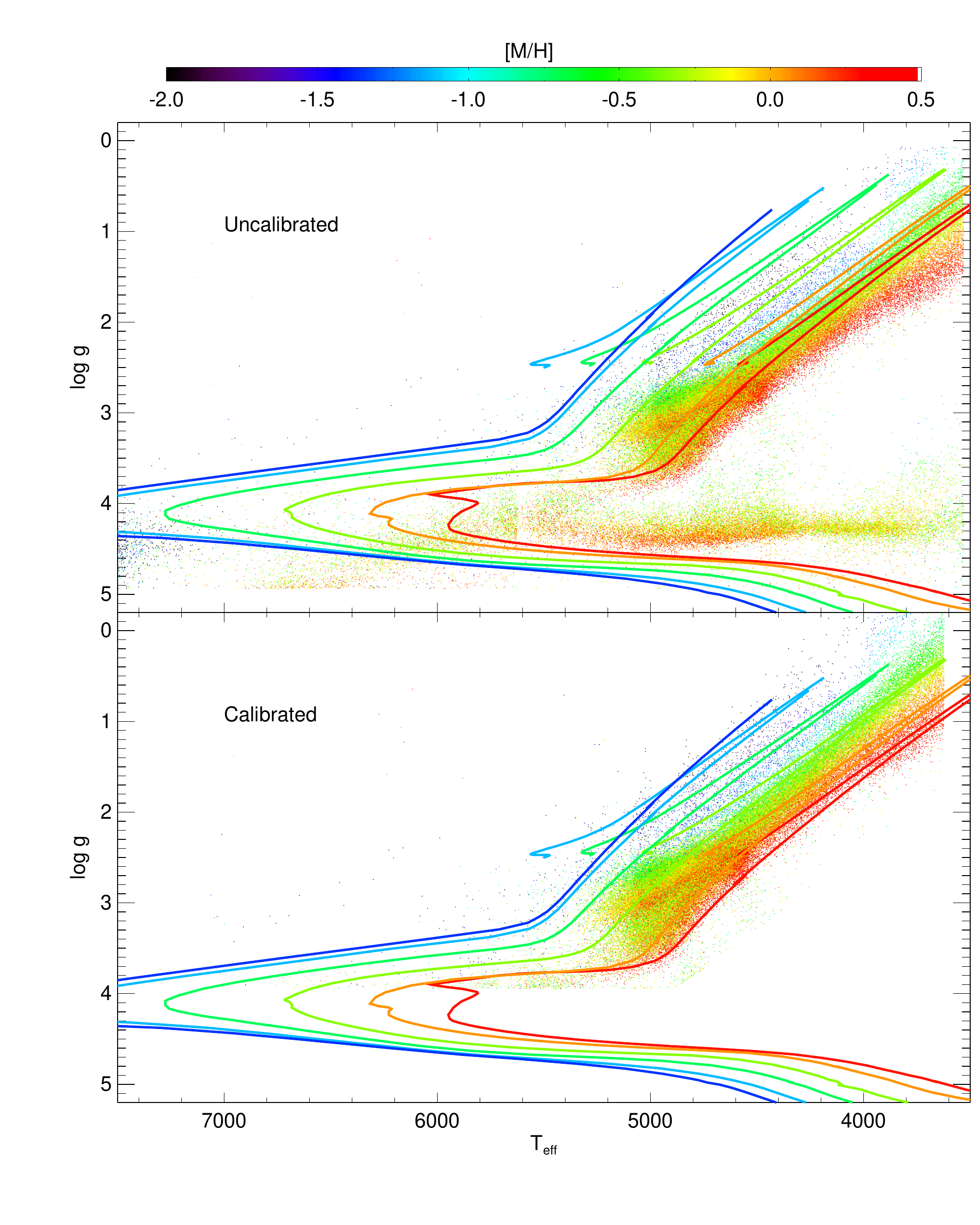}
\caption{Effective temperature - surface gravity - metallicity distributions for uncalibrated full APOGEE sample (top) and for calibrated giant sample (bottom). Isochrones from \citet{Bressan2012} for a 4 Gyr population are
overlaid.}
\label{fig:hr}
\end{figure*}

One way of testing the quality of the derived stellar parameters is to
see to what extent we recover the expected trends on an HR diagram. Figure
\ref{fig:hr} presents an HR diagram from both the uncalibrated APOGEE DR12
sample (top) and the calibrated sample (bottom), with points color-coded
by their derived metallicity, \mh. We also overlay isochrones for a 4
Gyr old population \citep{Bressan2012}.

The uncalibrated results (top panel) show that the dwarfs, while clearly
separated from the giants, have stellar parameters that are offset from the
stellar models.  In part, this is probably due to the exclusion of a
stellar rotation dimension in the synthetic grid, but since not all stars are
expected to have significant rotation, there are probably additional
effects as well.  While the ASPCAP parameters are of sufficient quality
to separate dwarfs from giants, these issues, along with the lack of
calibrators available for stars of high surface gravity, lead us to
recommend that the parameters and abundances for dwarfs be used only
with caution, and no attempt has been made to calibrate them.

The bottom panel of Figure \ref{fig:hr} shows the calibrated giant sample.
The calibrated parameters show the expected change in the locus of
points as a function of metallicity. This agreement, which is obtained
without imposing any assumptions about the relation between effective
temperature, surface gravity, and metallicity, is quite encouraging. We
note that some scatter in the locus at fixed metallicity is expected
if there exists a range of ages at given metallicity; this full sample
includes stars from across the Milky Way, so this is not expected to be
a coeval sample. In particular, the stars with lower metallicity may be
older than the 4 Gyr old isochrones shown here; older isochrones would
be in better agreement with the derived ASPCAP parameters.

The problems with the surface gravities for red clump stars are visible,
as stellar evolution models predict that red clump stars should
have near constant surface gravity, while the ASPCAP sequence slopes to
higher surface gravity at higher temperature. The ASPCAP gravities for
RC stars are also higher than expected from the stellar models (as was
also seen in the comparison to the asteroseismic gravities in \S \ref{sect:logg}).

\begin{figure}[h]
\includegraphics[width=0.5 \textwidth]{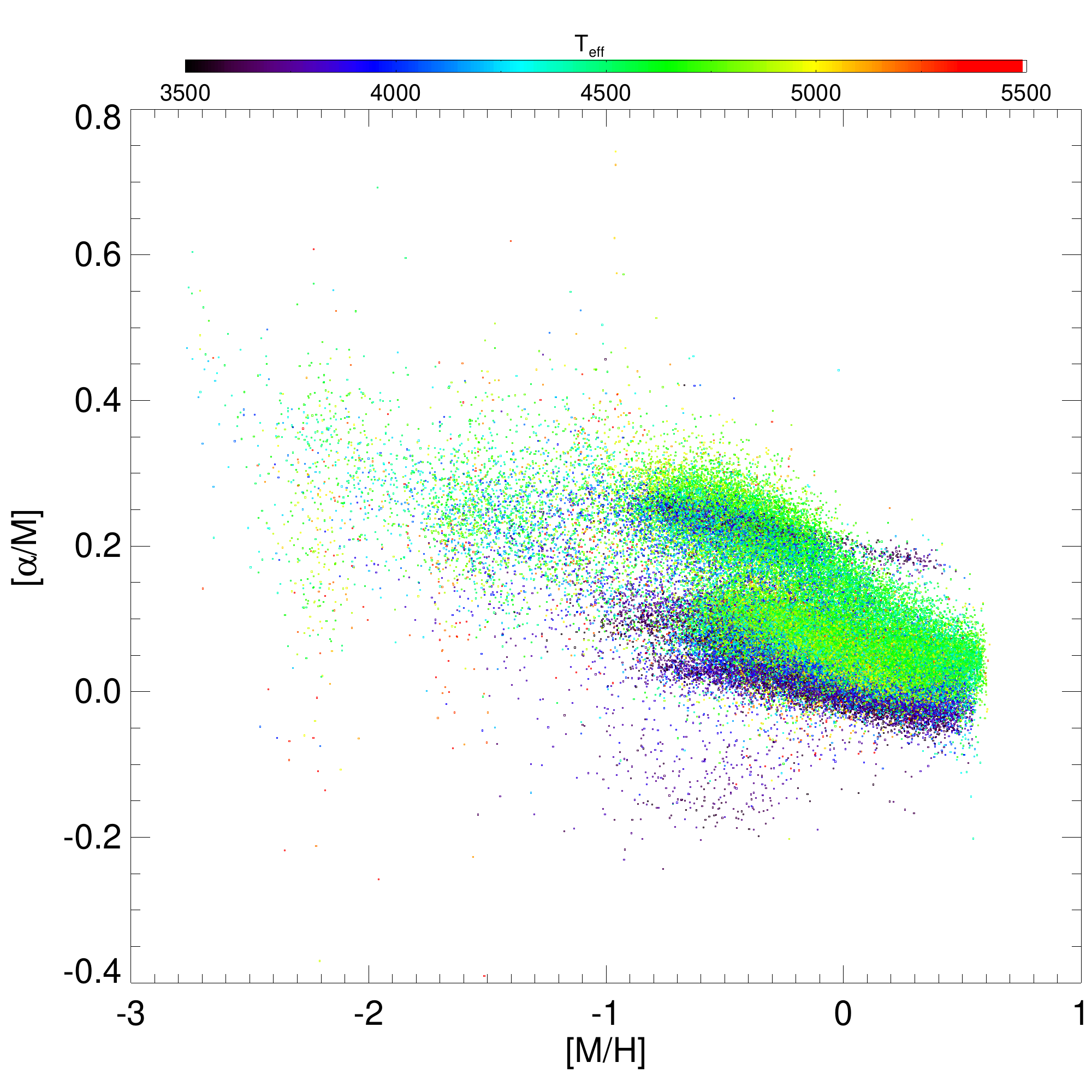}
\caption{Metallicity - \am\  - temperature relation for calibrated APOGEE giant sample.}
\label{fig:alphafe}
\end{figure}

Figure \ref{fig:alphafe} shows the \am\  vs \mh\  relation for 
the calibrated giant sample, color coded by temperature. The expected
trends towards higher \am\  at lower metallicity is apparent. However,
the cooler stars (blue points) appear to deviate from the
locus of stars of other temperatures. While it is possible that there
are some astrophysical effects contributing here (cooler stars are more
likely to be at larger distances), a more likely interpretation is that
 the \am\ results (and probably other
abundances) are less reliable at the cooler end of the ASPCAP temperature range,
i.e. $T<4000$ K. This is also the temperature range where the quality of
the fits gets significantly worse. As a result, users should exercise 
caution when using results for the cooler stars.

\subsubsection{Stellar abundances}

\begin{figure*}[h]
\includegraphics[width=\textwidth]{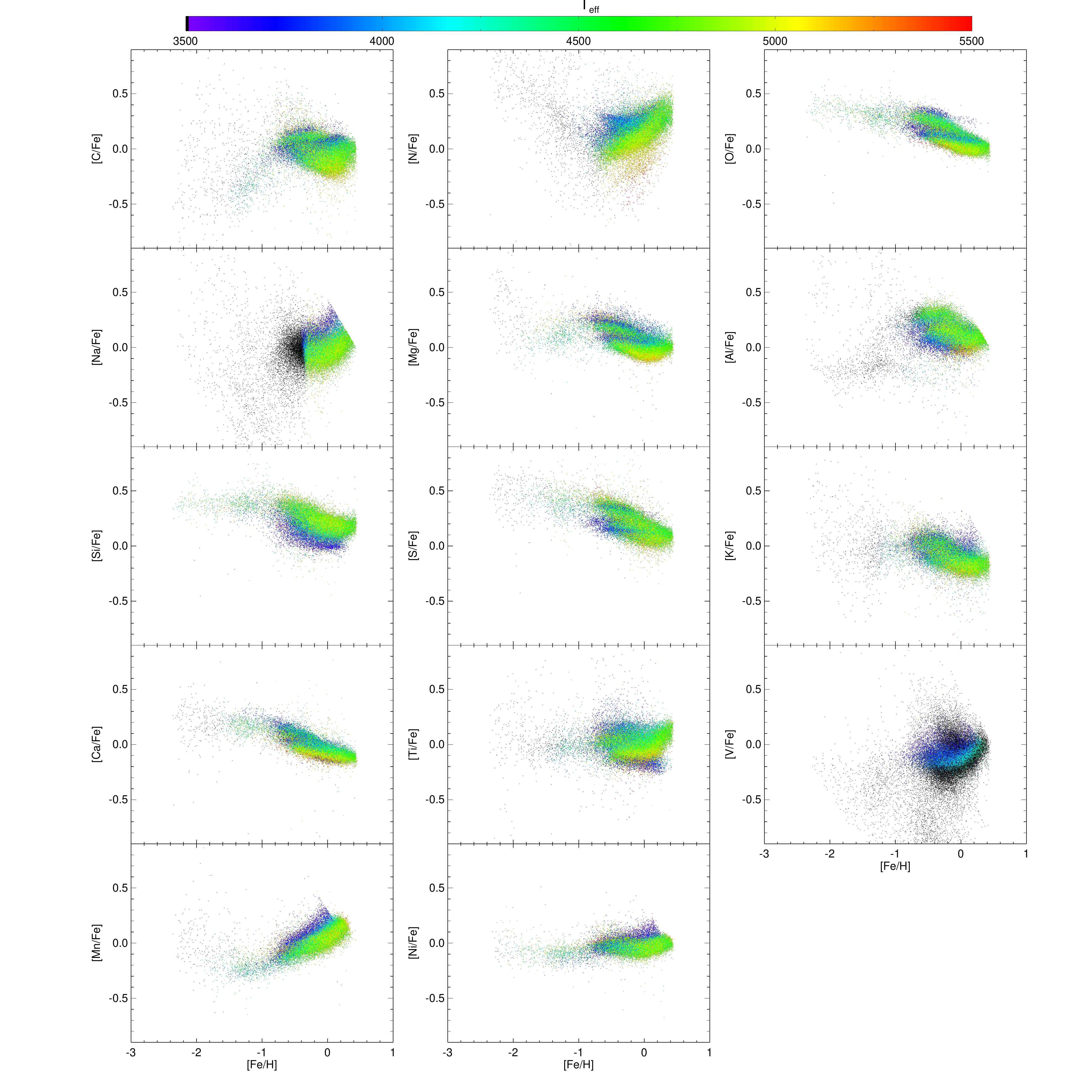}
\caption{Abundance ratio plots for high S/N ($>$200) stars from full APOGEE giant sample.
Points are color coded by effective temperature; black points are stars
with large uncertainties (empirical uncertainties greater than 0.1 dex).}
\label{fig:abundances}
\end{figure*}

Figure \ref{fig:abundances} presents plots of \xfe\ vs. \feh\  for all
15 elements for a subsample of stars with high S/N ($>200$); the scatter
in these plots increases for lower S/N.  These are presented to assess
to what degree some of the expected trends are observed;
a more extensive discussion of these plots in terms of chemical evolution
is beyond the scope of this paper.  Points in the plots are color-coded
by effective temperature; black points are those for which the empirical
uncertainties (\S \ref{sect:errors}) are larger than 0.1 dex.

%
%
%

Previous studies from near the solar
neighborhood have shown, in general, enhanced [$\alpha$/Fe] ratios at
lower metallicities, while other elements tend to have less variation.
The APOGEE results for the $\alpha$ elements O, Mg, Si, S, and
Ca show trends of increasing abundance relative to Fe at lower
metallicity, as expected, although the behavior of Mg at the lowest
metallicities indicates some possible issues in that regime.  However,
Ti does not indicate any significant trend. Given that such a trend
has been clearly observed for titanium in the solar neighborhood (e.g.,
\citealt{Bensby2014}), this, along with the individual star comparison
discussed previously, suggests that there is some issue that may be
affecting the reliability of the ASPCAP Ti abundance.  Possible issues
with the ASPCAP derived Ti abundances will be investigated further in
future work.

While the shape of the $\alpha$ element locii are generally as expected,
the location of the suggest possible problems with the overall accuracy
(zeropoint) of the APOGEE/ASPCAP abundances. The Si and Su abundances
ratios ([Si/Fe] and [S/Fe]) are greater than zero at solar metallicity
by 0.1-0.2 dex, while the Ca abundance ([Ca/Fe]) is around -0.1. As noted
above, some offsets are perhaps not unexpected since we do not reference our abundances
to those of any well-calibrated object.

Results for Na and V show large scatter; features from these elements are
weak and most likely the scatter results from limited precision of the
APOGEE measurements.

The points in Figure \ref{fig:abundances} are color-coded by the
derived effective temperature, so that trends with temperature may
be discerned. In general, it seems that temperature trends are relatively
small, with the possible exception of those for Ti, Mn, and Ni.
However, given that cooler giants are more luminous, stars
of different temperatures likely sample different regions of the Milky
Way, so apparent abundance trends with temperature do
not necessarily represent an issue with the abundance determinations,
as the chemical relations may vary across the Milky Way.

Overall, these results support our previous assessment that the internal
precision of APOGEE/ASPCAP abundances is fairly good, while the external
accuracy is more uncertain.



\subsection{Persistence}
\label{sect:persist}

As noted in the description of the instrument, portions of two
of the SDSS-III/APOGEE detectors have strong persistence that
affects roughly one-third of the fibers over 40\% of the APOGEE spectral range.
Because the effect is complicated, not fully understood, 
and depends on the details of the exposure history, no attempt has
yet been made to correct for it in SDSS-III/APOGEE analysis.

We attempt to determine whether persistence has a significant effect
on the derived stellar parameters and abundances by comparing results for
objects whose spectra were recorded outside the persistence region
in all visits against results for objects whose spectra were recorded
within the highest persistence region in all visits; there are some
intermediate cases where some visits are recorded in the persistence
region, and some outside of it (since plates are generally replugged
between visits), as well as cases where spectra fall on regions of intermediate
or lower persistence.

\begin{figure}
\includegraphics[width=0.5 \textwidth]{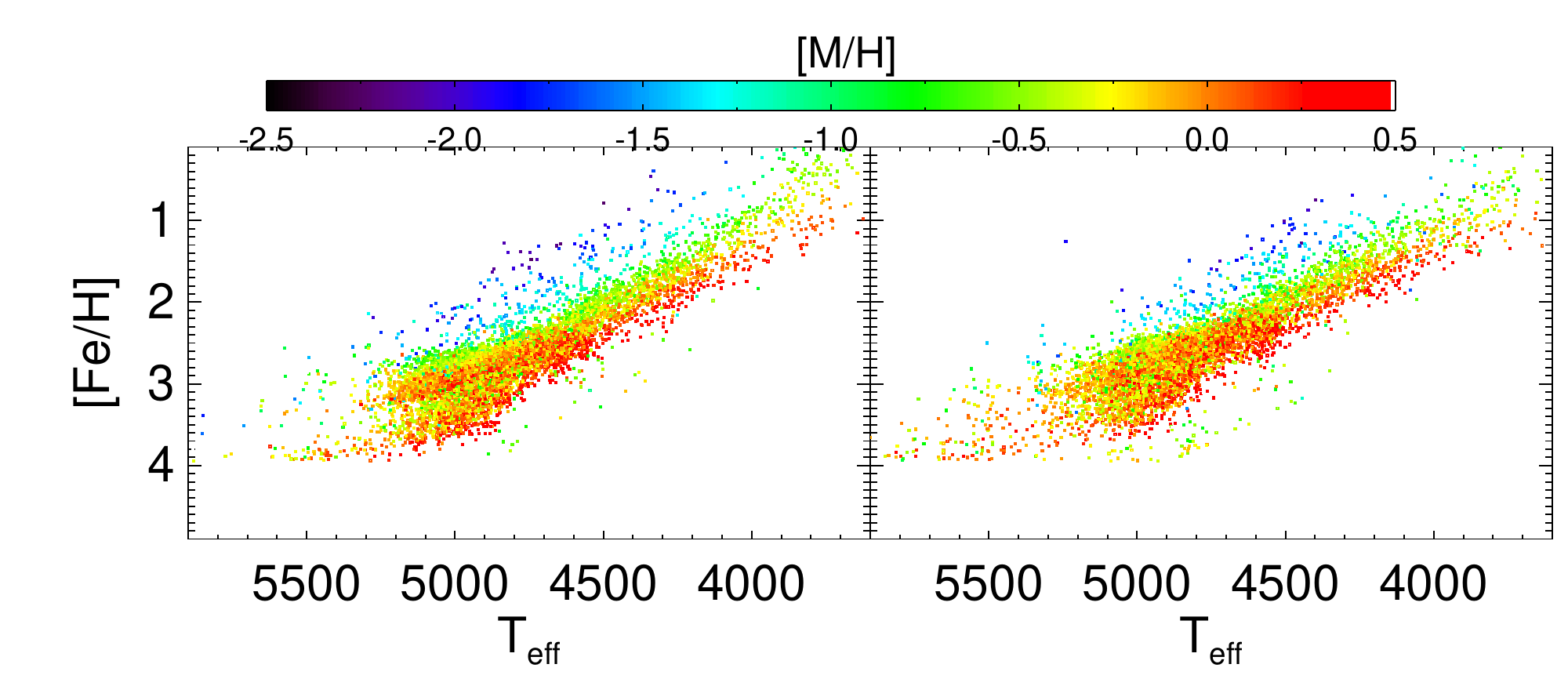}
\caption{Persistence comparison for \teff\  vs \logg for stars with $H>11$. 
Left panel has stars with all observations
outside of persistence region, while right panel has stars with all observations inside
of persistence region. Points are color-coded by temperature.}
\label{fig:persist_f_param}
\end{figure}

\begin{figure}
\includegraphics[width=0.5 \textwidth]{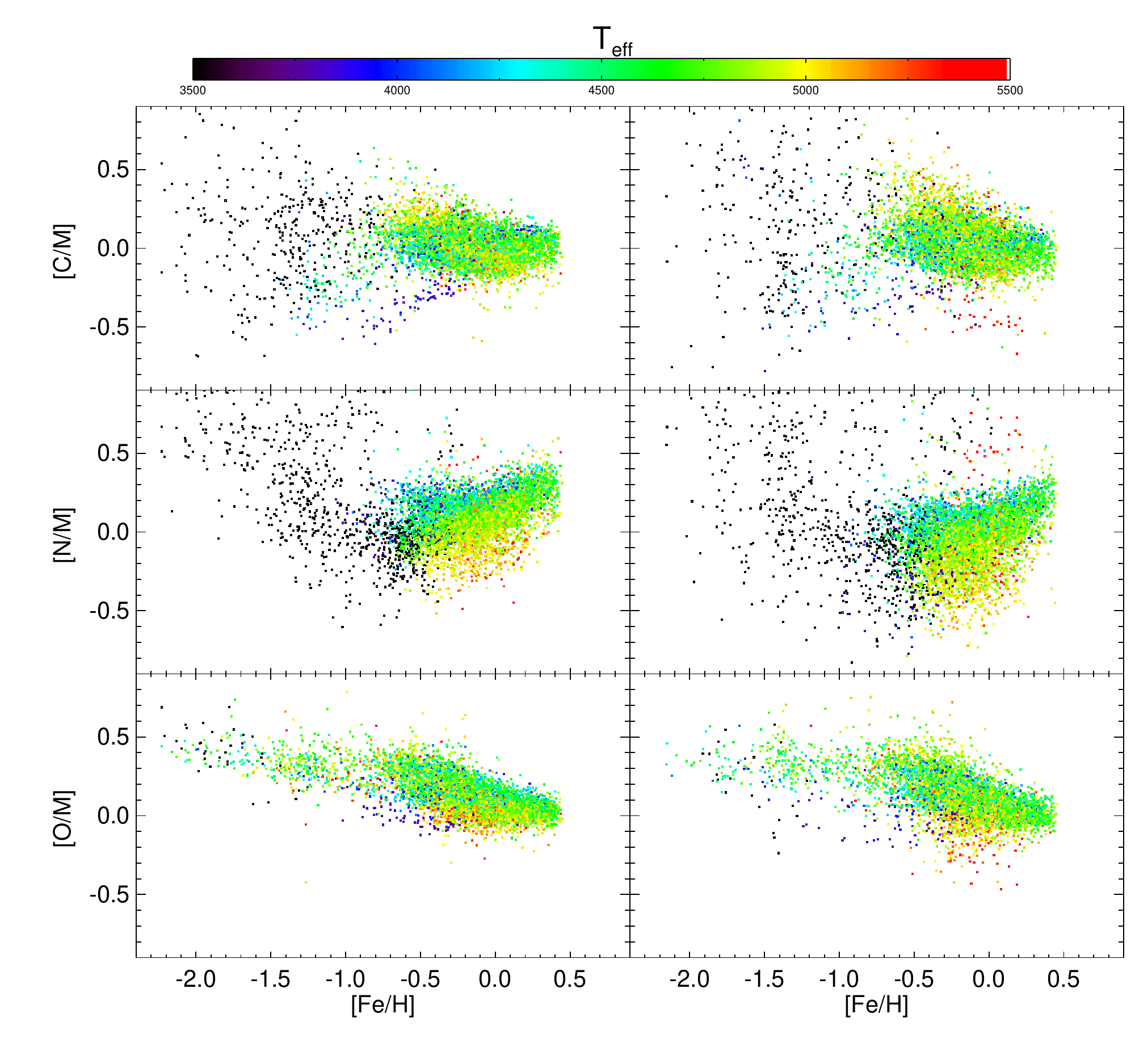}
\caption{Persistence comparison for CNO abundances for stars with $H>11$. 
Left panel has stars with all observations
outside of persistence region, while right panel has stars with all observations inside
of persistence region. Points are color-coded by temperature; black points have empirical
uncertainties greater than 0.1 dex.}
\label{fig:persist_f_cno}
\end{figure}

\begin{figure}
\includegraphics[width=0.5 \textwidth]{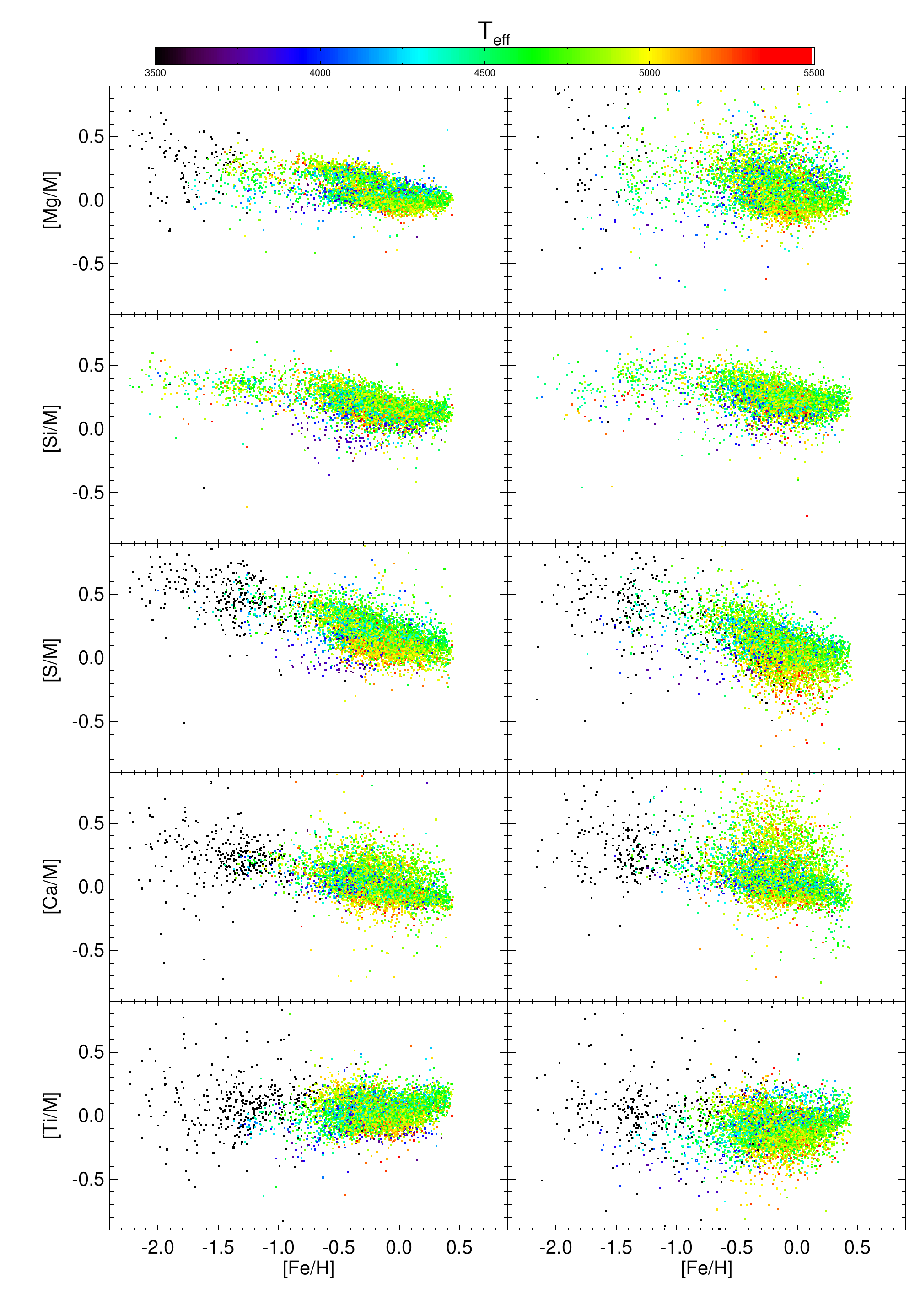}
\caption{As Figure \ref{fig:persist_f_cno}, but for $\alpha$-element abundances (Mg, Si, S, Ca, Ti). }
\label{fig:persist_f_alpha}
\end{figure}

\begin{figure}
\includegraphics[width=0.5 \textwidth]{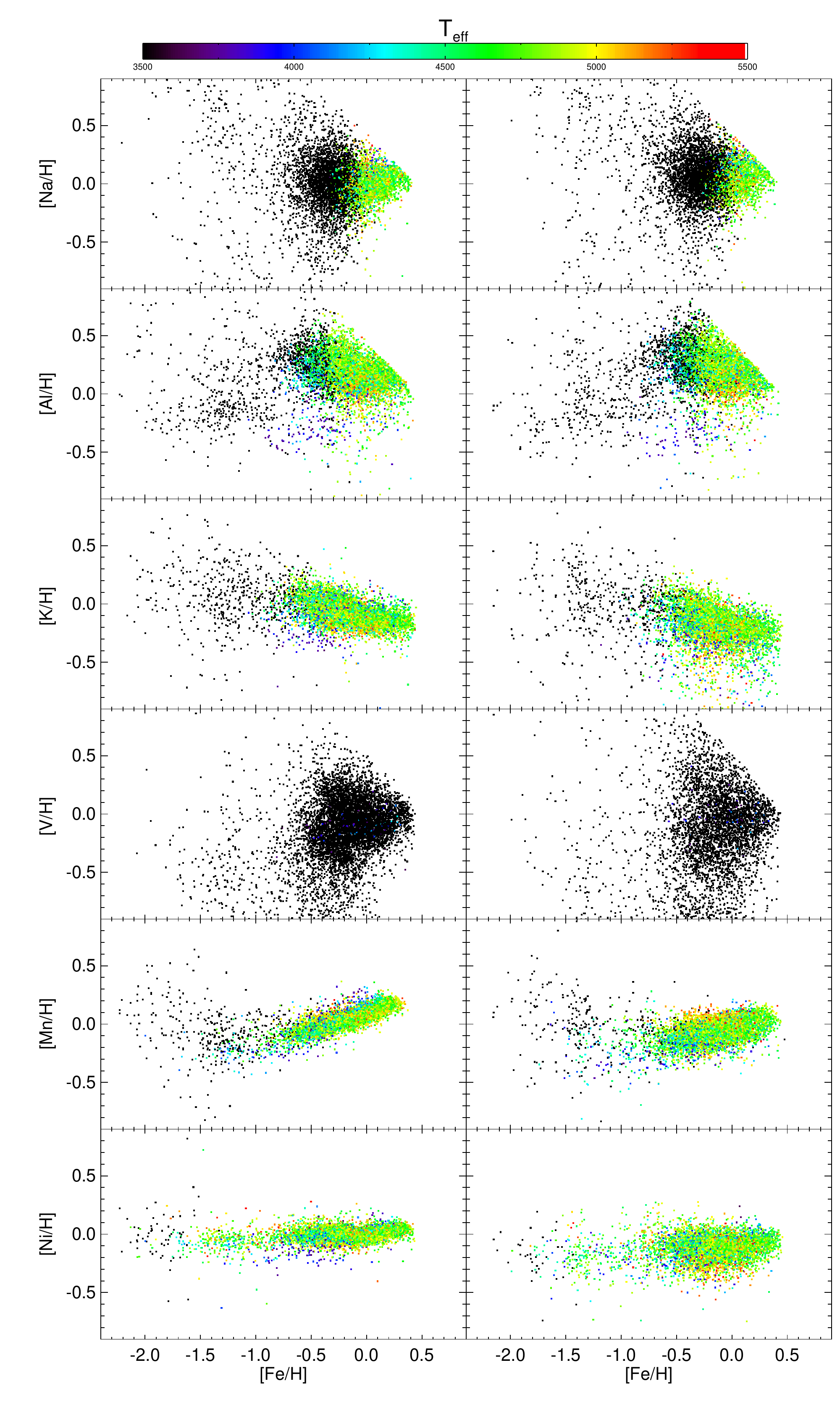}
\caption{As Figure \ref{fig:persist_f_cno}, but for Na, Al, K, V, Mn, and Ni. }
\label{fig:persist_f_fe}
\end{figure}

Results are shown in Figure \ref{fig:persist_f_param}, \ref{fig:persist_f_cno},
\ref{fig:persist_f_alpha}, and \ref{fig:persist_f_fe}. In all cases, we
show only fainter stars with $H>11$, which tend to be more affected by
persistence than brighter stars. The left plots show the results for
stars that do not fall in the persistence region for \textit{any} of the
visits, and the right plots show results for stars that fall in the high
persistence region in \textit{all} visits. Figure \ref{fig:persist_f_param}
suggests that persistence does not seem to impact the parameters
dramatically.  However, for the individual elemental abundances, persistence
can have a noticeable effect. In particular, the relations for 
N, Mg, S, K, Ca, Ti, Mn, and Ni show significantly larger scatter or 
offsets for stars that fall in the high persistence region; given that
some of these have features that are not within the persistence region,
the stellar parameters must be affected at some level.

As a result, users are cautioned about the use of some of the individual
element abundances for stars that were observed within the persistence
region. These stars can be identified using the star quality bitmask
described below.

\section{Data products and access}
\label{sect:access}

APOGEE data can be accessed in several different ways.

\subsection{Catalog Archive Server (CAS)}

Catalog data are also stored in a database called the
Catalog Archive Server (CAS), which can be accessed at
\url{http://skyserver.sdss3.org/dr12}. The CAS contains a number of ways of
interacting with the database, including low-level SQL access through the
CasJobs interface at \url{skyserver.sdss3.org/Casjobs}.  More information
is available at \url{skyserver.sdss3.org}.

The CAS provides access to several database tables:

    \begin{itemize}
      \item apogeeStar --- a table with basic information for each combined
spectrum, including targeting flags, radial velocity information, and 
the \flag{STARFLAG} and \flag{ANDFLAG} data quality bitmasks for each 
individual star;

      \item aspcapStar --- a table with the ASPCAP parameters, abundances, 
uncertainties, and bitmasks flagging possible conditions with the  
parameters and abundances;

      \item apogeeVisit --- a table with the radial velocity information for
each individual visit of each star;

      \item apogeeObject ---  a table with all of the ancillary catalog/targeting
information for the parent sample of stars in each field from which the observed
targets were drawn;

      \item apogeePlate --- a table with information about each plate observed;

      \item apogeeDesign --- a table that contains information about each plate
design; 

      \item apogeeField --- a table with information about each field that was 
observed, which includes the \flag{LOCATION\_ID} numeric value that is used
to index different fields, and also the planned number of visits for each
field.

    \end{itemize}

There are also some tables that link these tables, e.g., apogeeStar to apogeeVisit.

This database only contains catalog information derived from the
spectra, but not the spectra themselves.

\subsection{Science Archive Server (SAS)}

The data files created at multiple stages in the data reduction
and processing pipeline are available on the Science Archive Server (SAS), which
is accessible through the web at \url{http://data.sdss3.org/sas/dr12}. 

\subsubsection{FITS tables}

Summary FITS tables are available that present compilations of data for
all stars in the data release. These FITS tables are the original data
source for the CAS database tables:

   \begin{itemize} 

    \item allStar --- summary information on each individual star observed in
SDSS-III/APOGEE, which includes mean barycentric radial velocity, ASPCAP
parameters and abundances as derived from the summed spectra, and a
compilation of ancillary targeting data (for DR12, allStar-v603.fits);

    \item allVisit --- summary information on each individual visit of each star
in SDSS-III/APOGEE, which includes the barycentric radial velocity of
each visit, along with a compilation of ancillary targeting data (for
DR12, allVisit-v603.fits).

    \item allPlates --- summary  information for all observed plates,
include the location and plate design information (for DR12, allPlates-v603.fits).
   \end{itemize}

Both allStar and allVisit files are FITS files with several header/data
units (HDUs). The first HDU consists of the main data, with an entry for
each star (allStar HDU1) or visit (allVisit HDU1). The objects are sorted
by increasing right ascension.  The second HDU in both files contains
a 360 element integer array giving the HDU1 array index of the first
object larger than each degree of right ascension, which may be useful
for a rapid search through the table for an object of known RA. The
third HDU contains information about the order of array values in the
\flag{PARAM} (\flag{PARAM\_SYMBOL}) and \flag{ELEM} (\flag{ELEM\_SYMBOL}
and \flag{ELEM\_VALUE})
arrays in HDU1; these are constant for all objects and thus do not need
to be repeated for each object.

The allStar file provides pointers into the allVisit file to identify the
visits that went into the combined spectra.  

There are also several summary targeting FITS tables:
   \begin{itemize}

    \item apogeeDesign gives targeting information for each plate design

    \item apogeePlate gives targeting information for each plate 
(design\_id, location\_id, hour angle, temperature, epoch)

    \item apogeeField gives targeting information for each field (name,
location\_id, coordinates, and expected number of visits)

    \item apogeeObject gives all of the ancillary catalog/targeting
information for \textit{all} stars in each field from which the observed
targets were drawn

   \end{itemize}
Targeting terms are defined in \citet{Zasowski2013}, particularly in the
glossary of that paper.


\subsubsection{Spectral data}
\label{sect:spectra}

The spectral data files created at multiple stages in the data reduction
and processing pipeline are also available on SAS.
Some of the data that might be of the most interest include:

   \begin{itemize}
     \item \textit{apVisit} files --- FITS image files with individual visit spectra of
each star on each MJD in which it was observed;

     \item \textit{apStar} files --- FITS image files with combined
(across multiple visits) spectra of each star observed with APOGEE, along with
uncertainty and mask arrays, as well as the individual visit spectra resampled
to the apStar rest wavelength scale;

     \item \textit{aspcapStar} files --- FITS image files for each star that give
the continuum normalized spectra from which parameters and abundances
are derived, the adopted uncertainties in the spectra, and the 
best-matching synthetic spectra, along with derived parameters and
abundances in the image header;

     \item \textit{apField} files --- FITS tables for each field that include 
information and the radial velocities for all stars in the field;

     \item \textit{aspcapField} files --- FITS tables for each field that include
the derived ASPCAP parameters and abundances for all stars in the field,
along with the spectra and best-matching synthetic spectra.

   \end{itemize}
These files are described in more detail in \S \ref{sect:usingspectra}. Note
that the information in the apField and aspcapField files is compiled including
all fields in the allStar file discussed above.

In addition to these main products, the SAS contains \textit{all} of the
data from APOGEE, from the raw data cubes to the final processed
spectra and derived parameters, including all of the intermediate data products. 

Table \ref{tab:directory} provides a rough guide to the overall data
structure in the SAS.  All of these data 
products are documented through the SDSS datamodel,
available at \url{http://data.sdss3.org/datamodel}

\begin{deluxetable*}{ll}[h]
\tablecaption{SAS directory structure}
\tablewidth{\textwidth}
\tablehead{
\colhead{Directory}&\colhead{contents}
}
\startdata
APRED\_VERS & top level directory for a given visit reduction version\\
APRED\_VERS/red/MJD & reduced data for individual exposures, \\
                    & \ \ ap2D and ap1D\\
APRED\_VERS/apo25m/PLATE/MJD & reduced visit data for 2.5m data: apVisit files\\
APRED\_VERS/apo1m/PLATE/MJD &reduced visit data for 1m data: apVisit files\\
APRED\_VERS/APSTAR\_VERS & top level for a given star combination version\\
APRED\_VERS/APSTAR\_VERS/apo25m/LOCATION\_ID & combined, resampled spectra: apStar files for 2.5m data\\
APRED\_VERS/APSTAR\_VERS/apo1m/PROGRAM\_ID & combined, resampled spectra: apStar files for 1m data\\
APRED\_VERS/APSTAR\_VERS/ASPCAP\_VERS & top level directory for a given ASPCAP \\
                 & \ \ version/configuration\\
APRED\_VERS/APSTAR\_VERS/ASPCAP\_VERS/location\_id & FERRE files for a given location\\
APRED\_VERS/APSTAR\_VERS/ASPCAP\_VERS/RESULTS\_VERS & top level directory for a given RESULTS (calibration)  \\
  & version: allStar and allVisit summary tables\\
APRED\_VERS/APSTAR\_VERS/ASPCAP\_VERS/RESULTS\_VERS/location\_id & ASPCAP results in FITS files for  given field: \\
  & aspcapStar and aspcapField files\\
\enddata
\label{tab:directory}
\end{deluxetable*}

\subsection{Web interface to the SAS}

A web application  provides an interface to the SAS through a
front-end database called the SAS database (SASDB). This provides a tool by
which individual spectra can be interactively inspected without having to download the
data, and also provides an interface by which you can search for data to
download either for individual objects, or for a bulk search of multiple
objects. This can be found at \url{http://data.sdss3.org}

\section{Using APOGEE data}
\label{sect:usingdata}

\subsection{Bitmasks, data quality, and target classes}

SDSS uses bitmasks to flag stars with quality issues that may
affect the accuracy of the derived parameters and abundances, as 
well as to provide information on targeting.  A bitmask
is an integer in which individual bits (corresponding to powers of two
in a decimal representation) can be set to indicate a variety of conditions.
Users are advised to routinely check the contents of these. 

\subsubsection{Targeting bitmasks}

There are different target classes for objects that appear in
APOGEE data releases.  While the bulk of the targets were chosen
for the ``main" survey (i.e., stars selected with a simple color selection),
there are a number of other target
classes, as described in detail in \citet{Zasowski2013}. These include
special targets such as the cluster targets used for the calibration,
targets for ancillary science programs, and observations obtained
with the NMSU 1m feed to the APOGEE spectrograph. The targeting
class for each target is documented in a pair of targeting bitmasks,
\flag{APOGEE\_TARGET1} and \flag{APOGEE\_TARGET2}, where the bit
definitions are defined in \citet{Zasowski2013} and documented at
\url{http://www.sdss.org/dr12/algorithms/bitmasks/#APOGEE_TARGET1} and
\url{http://www.sdss.org/dr12/algorithms/bitmasks#APOGEE_TARGET2}. We
provide character ``translations" of the targeting bitmasks in
the \flag{APOGEE\_TARGFLAGS} column in summary data files.

Note that the main survey targets are \textit{not} defined by a single bit
in the targeting bitmasks; instead, they can be identified as targets
that have one of the \flag{APOGEE\_SHORT}, \flag{APOGEE\_INTERMEDIATE},
or \flag{APOGEE\_LONG} bits set. For convenience, we have included a
separate bitmask, \flag{EXTRATARG}, in the summary files and in the CAS
tables. This bitmask has a value of zero for all main survey targets,
but has different bits set if the target is either a star selected for
science outside of the main survey (bit 0), commissioning data (bit 1),
a telluric star (bit 2), or a star observed with the NMSU 1m (bit 3),
see \url{http://www.sdss.org/dr12/algorithms/bitmasks#APOGEE_EXTRATARG}.

There are some objects that may appear more than once in the summary
files or the CAS tables. Many of these are stars that were
observed during the commissioning period. These observations are kept
separate because the quality of the the data is poorer (worse LSF).
Since these data are not homogeneous with the main survey data, they
are not analyzed with ASPCAP; they are included because they still 
provide useful radial velocity information and may reveal spectral
variations over time (e.g., double-line spectroscopic binaries or
emission line stars).  Apart from this,
there are $\sim$ 1600 stars that appear more than once with survey
data. This occurs because these objects were targeted in different, but
overlapping fields. In some cases, this was intentional (in particular,
for some of the ancillary science programs), but for others the targets
were inadvertently selected independently in two different overlapping fields.
For all duplicate targets, we choose a ``primary" observation as the
observation with the highest S/N, and all other observations of this
target have a bit (bit 4) set in the \flag{EXTRATARG} bitmask.

The most homogeneous sample are those stars with \flag{EXTRATARG}=0.
This will give only main survey targets, excluding all special targets,
commissioning data, 1m objects, telluric stars, and duplicate targets.

\subsubsection{STARFLAG bitmask}

\begin{deluxetable*}{lll}[h]
\tablecaption{STARFLAG bitmask}
\tablewidth{\textwidth}
\tablehead{
\colhead{Condition}&\colhead{Bit}&\colhead{Description}
}
\startdata
BAD\_PIXELS&           0&Spectrum has many bad pixels ($>$40\%):  BAD\\
COMMISSIONING&         1&Commissioning data (MJD$<$55761), non-standard configuration, poor LSF: WARN\\
BRIGHT\_NEIGHBOR&      2&Spectrum has neighbor on detector more than 10 times brighter: WARN\\
VERY\_BRIGHT\_NEIGHBOR&3&Spectrum has neighbor on detector more than 100 times brighter: BAD\\
LOW\_SNR&              4&Spectrum has low S/N (S/N$<$5): BAD\\
&                    5-8& Currently unused\\
PERSIST\_HIGH&         9&Spectrum has significant number ($>$20\%) of pixels in high persistence region: WARN\\
PERSIST\_MED&         10&Spectrum has significant number ($>$20\%) of pixels in medium persistence region: WARN\\
PERSIST\_LOW&         11&Spectrum has significant number ($>$20\%) of pixels in low persistence region: WARN\\
PERSIST\_JUMP\_POS&   12&Spectrum shows obvious positive jump in blue chip: WARN\\
PERSIST\_JUMP\_NEG&   13&Spectrum shows obvious negative jump in blue chip: WARN\\
&                  14-15&Currently unused\\
SUSPECT\_RV\_COMBINATION&          16&WARNING: RVs from synthetic template differ significantly \\ 
  &  & from those from combined template\\
SUSPECT\_BROAD\_LINES&17&WARNING: cross-correlation peak with template significantly \\
  &  &broader than autocorrelation of template\\
&               18-31&Currently unused\\
\enddata
\label{tab:starflag}
\end{deluxetable*}

Each visit spectrum has an associated \flag{STARFLAG} bitmask, which
flags a number of conditions as given in Table \ref{tab:starflag} and
at \url{http://www.sdss.org/algorithms/dr12/bitmasks/#APOGEE_STARFLAG}.
If \flag{BAD\_PIXELS}, \flag{VERY\_BRIGHT\_NEIGHBOR}, or \flag{LOW\_SNR}
is set for a visit spectrum, that spectrum is deemed bad and not used
to obtain the final combined spectrum.
The 
\flag{PERSIST\_HIGH}, \flag{PERSIST\_MED}, and \flag{PERSIST\_LOW} bits
flag stars that may be significantly affected by the persistence
on the blue detector; they are set if 20\% or more pixels are in the
persisence region, with the three different levels representing different
amplitude of the persistence affect (note that \flag{PERSIST\_LOW} is still
significant persistence: stars outside the persistence region altogether
do not have any of the these bits set).

The spectra that are combined from the individual visit spectra
have \flag{STARFLAG} set to the logical OR of the \flag{STARFLAG} values
of the individual visits. In addition, there is another ANDFLAG that is
set to the logical AND of the flags from the individual visits.

\begin{deluxetable*}{lll}[ht]
\tablecaption{ASPCAPFLAG bitmask}
\tablehead{
\colhead{Condition}&\colhead{Bit}&\colhead{Description}
}
\startdata
TEFF\_WARN 	& 0 & WARNING on effective temperature (see PARAMFLAG[0] for details)\\
LOGG\_WARN 	& 1 & WARNING on log g (see PARAMFLAG[1] for details)\\
VMICRO\_WARN 	& 2 & WARNING on vmicro (see PARAMFLAG[2] for details)\\
M\_H\_WARN 	& 3 & WARNING on \mh\  (see PARAMFLAG[3] for details)\\
ALPHA\_M\_WARN 	& 4 & WARNING on \am\   (see PARAMFLAG[4] for details)\\
C\_M\_WARN 	& 5 & WARNING on \cm\  (see PARAMFLAG[5] for details)\\
N\_M\_WARN 	& 6 & WARNING on \nm\  (see PARAMFLAG[6] for details)\\
STAR\_WARN 	& 7 & WARNING overall for star: set if any of TEFF, \\
  &  &LOGG, CHI2, COLORTE, ROTATION, SN warn are set\\
CHI2\_WARN 	& 8 & high $\chi^2$ ($>$ 2*median at ASPCAP temperature) (WARN)\\
COLORTE\_WARN 	& 9 & effective temperature more than 1000K from photometric \\
  &  &temperature for dereddened color (WARN)\\
ROTATION\_WARN 	& 10 & Spectrum has broad lines, with possible bad effects: ratio of FWHM of cross-correlation \\
   &  & of spectrum with best RV template to FWHM of auto-correlation of template $>$ 1.5 (WARN)\\
SN\_WARN 	& 11 & $S/N<70$ (WARN)\\
	& 12-15 & Not currently used\\
TEFF\_BAD 	& 16 & BAD effective temperature (see PARAMFLAG[0] for details)\\
LOGG\_BAD 	& 17 & BAD log g (see PARAMFLAG[1] for details)\\
VMICRO\_BAD 	& 18 & BAD vmicro (see PARAMFLAG[2] for details)\\
M\_H\_BAD 	& 19 & BAD \mh\  (see PARAMFLAG[3] for details)\\
ALPHA\_M\_BAD 	& 20 & BAD \am\   (see PARAMFLAG[4] for details)\\
C\_M\_BAD 	& 21 & BAD \cm\  (see PARAMFLAG[5] for details)\\
N\_M\_BAD 	& 22 & BAD \nm\  (see PARAMFLAG[6] for details)\\
STAR\_BAD 	& 23 & BAD overall for star: set if any of TEFF, LOGG, CHI2, COLORTE, ROTATION, \\
   &  & SN error are set, or any parameter is near grid edge (GRIDEDGE\_BAD is set in any PARAMFLAG)\\
CHI2\_BAD 	& 24 & high $\chi^2$ ($>$ 5*median at ASPCAP temperature) (BAD)\\
COLORTE\_BAD 	& 25 & effective temperature more than 2000K from photometric temperature for dereddened color (BAD)\\
ROTATION\_BAD 	& 26 & Spectrum has broad lines, with possible bad effects: ratio of FWHM of cross-correlation \\
   &  & of spectrum with best RV template to FWHM to auto-correlation of template $>$ 2 (BAD)\\
SN\_BAD 	& 27 & $S/N<50$ (BAD)\\
	& 28-30 & Not currently used\\
NO\_ASPCAP\_RESULT 	& 31 & True for all commissioning data, and for spectra that are marked as bad\\
\enddata
\label{tab:aspcapflag}
\end{deluxetable*}

\begin{deluxetable*}{lll}
\tablecaption{PARAMFLAG bitmask}
\tablehead{
\colhead{Condition}&\colhead{Bit}&\colhead{Description}
}
\startdata
GRIDEDGE\_BAD 	& 0 &	Parameter within 1/8 grid spacing of grid edge\\
CALRANGE\_BAD 	& 1 &	Parameter outside valid range of calibration determination\\
OTHER\_BAD 	& 2 &	Other error condition\\
	& 3-7 &	Not currently used\\
GRIDEDGE\_WARN 	& 8 &	Parameter within 1/2 grid spacing of grid edge\\
CALRANGE\_WARN 	& 9 &	Parameter in possibly unreliable range of calibration determination\\
OTHER\_WARN 	& 10 &	Other warning condition\\
	& 11-15 & Not currently used	\\
PARAM\_FIXED 	& 16 &	Parameter set at fixed value, not fit\\
	& 17-31  & Not currently used\\
\enddata
\label{tab:paramflag}
\end{deluxetable*}

\subsubsection{ASPCAP bitmasks}
\label{sect:aspcapbitmasks}

Several bitmasks are used in the ASPCAP processing. There is
an overall \flag{ASPCAPFLAG}, which flags various conditions
as listed in Table \ref{tab:aspcapflag}. In particular, bits
are set if any of the parameters are near the spectral library 
grid edges, if the ASPCAP temperature differs significantly
from that expected from the observed colors, if the spectrum
shows signs of significant stellar rotation, or if the spectrum 
does not meet the survey requirement for $S/N$.

Of particular note is the \flag{STAR\_BAD} bit, which is set if any
of other bits (\flag{TEFF\_BAD}, \flag{LOGG\_BAD}, \flag{CHI2\_BAD},
\flag{COLORTE\_BAD}, \flag{ROTATION\_BAD}, \flag{SN\_BAD}) to
denote that the ASPCAP results are very likely not to be reliable.
Stars with this bit set should not be used for parameter/abundance
analysis. Note that some bits in \flag{ASPCAPFLAG} are informational/warning
only, so restricting a sample to \flag{ASPCAPFLAG}=0 is \textbf{not}
an appropriate filter to get good data.

In addition, there are \flag{PARAMFLAG} and \flag{ELEMFLAG} arrays for
each of the stellar parameters and individual element abundances that
provide additional information about why a given parameter may be denoted
\flag{BAD} in \flag{ASPCAPFLAG} or that flag potential issues with the
abundance determination for individual elements. Table \ref{tab:paramflag}
gives the definition of the \flag{PARAMFLAG} bits.

\subsection{Radial velocities}

Several measurements of radial velocities are provided, as discussed in the
following subsections. 

\subsubsection{Systemic radial velocities}

The average systemic velocity for each APOGEE star is given in
\flag{VHELIO\_AVG}; estimated uncertainties are given in \flag{VERR}
(weighted error), and \flag{VERR\_MED } (median error of individual
visit RVs).  The scatter obtained from multiple visits (with number of
visits given in \flag{NVISITS}) is given in \flag{VSCATTER}. Note that
objects with \flag{VSCATTER} $>$ 0.5 km/s are potentially RV-variable
objects, based on the typical RV accuracy (see \citealt{Nidever2015}).  
These velocities were determined by deriving relative visit RVs
by iterative cross-correlation of each visit against the combined spectrum
for the object, with an absolute scale established after convergence by
cross-correlation of combined spectra with the best-matching synthetic
spectrum in a RV template grid. This has the advantage of having no
template mismatch for the relative RVs (although a disadvantage is
that the combined template is not noiseless).  The parameters of the
best-matching grid point are given in \flag{RV\_TEFF}, \flag{RV\_LOGG},
and RV\_FEH, but these are not to be confused with the much better
determined ASPCAP parameters; no interpolation is performed in the
RV grid.

An alternative systemic velocity is given in \flag{SYNTHVHELIO\_AVG},
with uncertainties in \flag{SYNTHVERR} and \flag{SYNTHVERR\_MED} and
scatter in \flag{SYNTHVSCATTER}. These are determined by cross-correlation
of each visit with the RV template grid, and then subsequent combination.

The scatter between the two different systemic RV measurements
is given in \flag{SYNTHSCATTER}. Larger values here suggest
\textit{possible} issues with the derived RVs. For such objects, the
\flag{SUSPECT\_RV\_COMBINATION} bit in the \flag{STARFLAG} bitmask is set.

\subsubsection{Individual visit radial velocities}

The individual visit radial velocities are stored in the visit tables
(\textit{allVisit} FITS table in the SAS, and the \textit{allVisit} 
table in the CAS).
The visit RV information derived from cross-correlating each visit
against the combined spectra are stored with labels \flag{VREL} (the
measured RV, uncorrected for Earth motion), \flag{VRELERR} (uncertainty in
\flag{VERR}), and \flag{VHELIO} (measured RV corrected for barycentric
motion; we note the slight misnomer in the \flag{VHELIO} label since these are strictly
barycentric, not heliocentric velocities). The visit RVs derived from
cross-correlating each visit against the best-matching template are
stored in \flag{SYNTHVREL}, \flag{SYNTHVRELERR}, and \flag{SYNTHVHELIO}.

An initial estimate of the radial velocity for each for each visit is
made by cross-correlation of the individual visit spectra with the same coarse
grid of synthetic templates mentioned above. These are stored in the visit tables as
\flag{ESTVREL}, \flag{ESTVRELERR}, and \flag{ESTVHELIO}.

\subsection{Stellar parameters}

As described previously, each APOGEE spectrum is matched to a spectral grid
covering several dimensions --- 
$T_{eff}$, \logg, $v_{micro}$, \mh, \cm, \nm, \am ---
to derive a set of stellar parameters that are subsequently used for the
individual element abundance fits.

Calibrations to \teff, \logg, \mh, and \am, as described above, are
applied to the raw FERRE output; for \mh, both internal (temperature)
and external calibration is applied. These quantities are stored in
the \flag{PARAM} array in the allStar file (HDU1); the order of the
parameters in the array is given in the \flag{PARAM\_SYMBOL} array in HDU3
of the allStar file.  In the CAS, the parameters are given in individual
columns \flag{PARAM\_TEFF}, \flag{PARAM\_LOGG}, \flag{PARAM\_LOGVMICRO},
\flag{PARAM\_M\_H}, \flag{PARAM\_C\_M}, \flag{PARAM\_N\_M}, and
\flag{PARAM\_ALPHA\_M} in the aspcapStar table.

The calibrated effective temperatures and surface gravities are duplicated
in columns TEFF and LOGG in both the \textit{allStar} file and CAS tables. These
are the recommended values for these quantities. The \textit{allStar} file also
duplicates the parameter-level \mh\  and \am\   in \flag{PARAM\_M\_H} and
\flag{PARAM\_ALPHA\_M}.

The raw, uncalibrated FERRE output is stored in an \flag{FPARAM} array
in the allStar file, and in the CAS in columns \flag{FPARAM\_TEFF},
\flag{FPARAM\_LOGG}, \flag{FPARAM\_M\_H}, \flag{FPARAM\_C\_M},
\flag{FPARAM\_N\_M}, and \flag{FPARAM\_ALPHA\_M} in the aspcapStar table.

There is  a distinction between the parameter-level
\mh\  and \am\   values  and the individual elemental abundances
discussed below. The parameter-level quantities are fits to the
entire spectrum, and might weight different elements differently over
a range of temperatures. However, they might also carry higher accuracy
as averages over multiple elements and more wavelengths. Furthermore,
the parameter-level \mh\  is the \textit{only} abundance that is
\textit{externally} calibrated, based on the literature cluster
metallicities discussed previously.


Stars with unreliable parameters are flagged by having the
\flag{STAR\_BAD} bit in the \flag{ASPCAPFLAG} bitmask set. If
this bit is set, then other bits in \flag{ASPCAPFLAG} can be used to
determine the reason for the measurements being flagged as unreliable. The
\flag{STAR\_WARN} bit can be consulted to determine if derived parameters
may be suspect based on less stringent criteria.

As discussed previously, while the ASPCAP results can easily discriminate
dwarfs from giants, the results for dwarfs are significantly less 
secure and lack an extensive calibration sample. As a result,
\textit{no calibratrated parameters are provided for dwarfs}.

Based on the quality of the fits and external comparisons, we also 
note that ASPCAP results at cooler temperatures,
\teff $<$ 4000 K may be more uncertain.

\subsection{Stellar abundances}

For the 15 individual elemental abundances, the internal calibration
(\ref{sect:intcal})
of abundance with temperature is applied, and the results are stored
in the \flag{ELEM} array in the allStar file (HDU1); the order of
the elements in the array is by atomic number and is given in the
\flag{ELEM\_SYMBOL} array in HDU3 of the allStar file. Since carbon,
nitrogen, and the $\alpha$ elements are fit by varying the \cm, \nm,
and \am\   dimensions in the spectral library, respectively, the abundances
from FERRE for these elements are relative to the overall scaled-solar
metal content, \mh, while the abundances for other elements are relative
to hydrogen since they are fit by varying \mh. These differences in the
array contents are documented in the \flag{ELEM\_VALUE} array of HDU3,
which explicitly gives what values are presented in the arrays.

For convenience, we convert all calibrated 
abundances into \xh{X}\  values (by adding
\mh\  to the carbon, nitrogen and $\alpha$ element abundances) and
store these in named quantities: \flag{C\_H}, \flag{N\_H}, \flag{O\_H},
\flag{Na\_H}, \flag{Mg\_H}, \flag{Al\_H}, \flag{Si\_H}, \flag{S\_H},
\flag{K\_H}, \flag{Ca\_H}, \flag{Ti\_H}, \flag{V\_H}, \flag{Mn\_H},
\flag{Fe\_H}, \flag{Ni\_H} in the allStar file and also as individual
columns in the aspcapStar table in the CAS.
As discussed in \S \ref{sect:extcal}, these abundances do not have
any external calibration applied to them.

The uncalibrated abundances are stored in an \flag{FELEM} array in
the allStar file, and in individual columns \flag{FELEM\_C\_M},
\flag{FELEM\_N\_M}, \flag{FELEM\_O\_M}, \flag{FELEM\_Na\_H},
\flag{FELEM\_Mg\_M}, \flag{FELEM\_Al\_H}, \flag{FELEM\_Si\_M},
\flag{FELEM\_S\_M}, \flag{FELEM\_K\_H}, \flag{FELEM\_Ca\_M},
\flag{FELEM\_Ti\_M}, \flag{FELEM\_V\_H}, \flag{FELEM\_Mn\_H},
\flag{FELEM\_Fe\_H}, \flag{FELEM\_Ni\_H} in the CAS table. Note
that these are \textit{not} all converted into \xh{X}; as raw
values, they preserve the way in which they were derived.

The \flag{ASPCAPFLAG} should be consulted to
confirm that there are no significant issues with the stellar parameters. 
In addition, before using
the abundance of any particular element, it is important to confirm that
the derived value is not near an abundance grid edge. This is achieved by
consulting the appropriate \flag{ELEMFLAG} bitmask, looking to see if the
\flag{GRIDEDGE\_BAD} bit (value within 1/8 of a grid spacing of edge)
or the \flag{GRIDEDGE\_WARN} bit (value within 1/2 of a grid spacing
of edge) is set. This is especially critical when looking at abundance
ratios, where the value of the ratio may not immediately indicate a
problem with one or both of the abundances.

Similarly to the stellar parameters, abundances for dwarfs are more
uncertain, and calibrated abundances for dwarfs are not provided.

The calibrated giant abundances have a effective temperature correction 
applied, as described in \S \ref{sect:abuncal}. Since this
metallicity-independent liner correction is perhaps unlikely to represent
the full behavior of the abundances, a sample that covers a wide temperature 
range may have larger possibility for temperature-dependent systematics
than one that is restricted to a smaller temperature range. In addition,
the calibration sample only extends down to \teff $\sim$ 3800 K, so
the behavior at cooler temperatures is less well known; the current
results apply the offset at 3800 K for all cooler stars, but set
the \flag{CALRANGE\_WARN} bit in the \flag{ELEMFLAG} bitmask.

Results for some elements may be less reliable than for others. In 
particular, the titanium abundances show large scatter compared with
previous measurements for a calibration subsample, and do not show the
expected behavior with overall metallicity. Other elements like sodium
and vanadium have relatively few and weak features, and thus have
larger uncertainties.

In addition, as noted in \S \ref{sect:persist}, individual element
abundances for objects whose spectra fell in the the persistence region of
the blue detector may be more uncertain. These objects can be identified
by those having one of the \flag{PERSIST\_HIGH}, \flag{PERSIST\_MED},
or \flag{PERSIST\_LOW} bits set in the \flag{ANDFLAG} bitmask (for
objects falling in persistence region in \textit{all} visits), or those
with one of the persistence bits set in the \flag{STARFLAG} bitmask
(for objects falling in persistence region in \textit{any} of the visits).

The empirical error estimates derived from simple fits to the observed
scatter of abundances in clusters as a function of temperature,
metallicity and signal-to-noise (see \S \ref{sect:intcal}) are
stored in the \flag{ELEM\_ERR} array (and correspondingly named columns
\flag{X\_H\_ERR} in the CAS).  Note that the scatter used for these
estimates is the scatter in the raw FERRE abundance measurements, i.e.,
in \xm{X} for C, N and the $\alpha$ elements, but in \xh{X} for Na, Al,
K, V, Mn, Fe, and Ni.  The \flag{FELEM\_ERR} array
(and corresponding column names in the CAS) gives the raw FERRE errors.

For uncertainties in abundance ratios, it is not necessarily appropriate
to add the uncertainties of the individual abundances in quadrature, as
the scatter may not all result from random errors. If there are systematic
errors contributing to the scatter, it is possible that different elements
will be offset in the same way, such that abundance ratios could be more
robust. Further analysis on this issue is planned for the future.

\subsection{Selection effects}

As with all surveys, there are selection effects in the APOGEE catalog. 
Several effects to consider are:
\begin{itemize}
\item main sample selection: only a fraction of the stars in each field
are selected for observation, with a well-defined distribution in magnitude
and color --- see \citet{Zasowski2013} for details.

\item ancillary and calibration targets: in addition to the main survey targets, there
are calibration and ancillary program targets that are not selected
by the usual algorithm. While the non-main survey targets can be
recognized from the bitmasks (e.g., \flag{EXTRATARG}=0), there is
a subtle effect that any targets that are part of an ancillary program
are rejected from being selected as a part of the main sample, so it
is possible that some object types are actually underrepresented in
the main sample. 

\item abundance biases at warmer temperatures: the sample selection generally 
uses a (dereddened) blue color cut of $(J-K)_0>0.5$. While this was
designed to include metal-poor luminous giants, it does bias against
lower metallicity stars on the lower giant branch. Furthermore, the
abundance measurements may become less certain at warmer temperatures
because features are weaker.

\item abundance biases at cooler temperatures: the current spectral grids
are limited to $T_{eff}>3500$ K and \logg $>$ 0, so no cooler and
lower surface gravity stars appear with good ASPCAP
parameters. This biases against the most metal-rich stars, in particular,
for the most luminous giants as well as bright AGB stars. 
Furthermore, the abundance measurements may
become less certain at cooler ($T_{eff}<4000$ K) temperatures because of the
complexity of the spectra and current limitations in our ability to model
it well.

\end{itemize}

\subsection{Caveats}

There are issues that are discovered in public data
releases after the point when things can be modified.
The online SDSS/APOGEE documentation attempts to track these at
\url{http://www.sdss3.org/dr12/irspec/caveats}. At the time of the data
release, known caveats include:

\begin{itemize}

\item Uncertainties given as \xh{X} are actually computed from 
\xm{X} scatter for C, N and $\alpha$ elements.

\item Character translations for target flags (in \flag{APOGEE\_TARGFLAGS}) are 
missing in the \textit{apVisit} and \textit{allVisit} files for four ancillary programs
(\flag{APOGEE\_RV\_MONITOR\_IC348}, \flag{APOGEE\_RV\_MONITOR\_KEPLER},
\flag{APOGEE\_GES\_CALIBRATE}, \flag{APOGEE\_BULGE\_RV\_VERIFY}); they
are correct in the \textit{apStar} and \textit{allStar} file

\item Some targets observed in multiple fields and appear as duplicates (as
flagged with the \flag{EXTRATARG} flag discussed previously).

\item Some targets may have been selected independently for different
programs within different visits to the same field. As a result, there
are stars constructed from multiple visits for which a target flag bit
may be set in the combined spectra, but not in all of the visit spectra
that were used to construct it.

\item The \flag{APOGEE\_ANCILLARY} bit is not set for \flag{APOGEE\_KEPLER\_SEISMO} and \flag{APOGEE\_KEPLER\_HOST} targets.

\item For a few objects, slightly incorrect object/star names were used during 
the reduction.

\end{itemize}

\section{Using APOGEE spectra}

\label{sect:usingspectra}

\subsection{Raw data}

Raw data files are saved as \textit{apR-[abc]-ID8.apz} frames, where [abc] refers
to each of the three chips, and the ID8 is an eight digit number that
uniquely identifies each APOGEE exposure. The first four digits give
the number of days since 2010 December 31, i.e. MJD-55562. The last four
digits are a running number of the frames taken during that day.

Data from the three detectors are stored in separate FITS files, with
the longest wavelengths at the lowest columns in chip \textit{a} and the shortest
wavelengths at the highest columns in chip \textit{c}. Each file is compressed
using the FPACK algorithm \citep{seaman2010}; the resulting images are given as
an initial image plus a set of differences between adjacent up-the-ramp
reads. We highlight the fact that some reconstruction is necessary to give 
the raw data cubes by giving the \textit{apR} files a .apz extension name 
(rather than .fits), even though they are valid FITS files.
See \citet{Nidever2015} for more details.

\subsection{Visit spectra}

Individual visit spectra of each star on each MJD in which it was observed
are saved in \textit{apVisit-VERS-PLATE-MJD-FIBER.fits} files.  VERS represents the
data release version of the reduction routines, which is r5 for the DR12
release. PLATE is the SDSS plate identification that corresponds to a physical
set of holes into which fibers are plugged. MJD represents the modified Julian
date of the observations, and FIBER represents which fiber was used to obtain
the spectrum.  The FIBER number is defined such that the spectrum at the
highest row numbers has FIBER=1, while those at the lowest row numbers have
FIBER=300 (see \citealt{Majewski2015}). 
The association of the fiber number with an object name is provided
by a ``plugmap" file; this association is used to populate header information
with an object name and coordinates.

The \textit{apVisit} spectra combine the multiple dithered exposures
taken on a given night. They are on the native APOGEE pixel scale,
but with half-pixel sampling made possible by the dithered exposures.
Multiple extensions give uncertainty in the spectra, a bitmask flag for
individual pixel status (\url{http://www.sdss.org/dr12/algorithms/bitmasks/#APOGEE_PIXMASK}), a wavelength array,
the sky and telluric spectra used to correct the object spectrum, and
some other ancillary information.  The content of the files is summarized
in Table \ref{tab:apvisit}.

\begin{deluxetable}{ll}[t]
\tablecaption{Contents of \textit{apVisit} files}
\tablehead{
\colhead{HDU}&\colhead{contents}
}
\startdata
0 & main header with target information, starflag\\
1 & flux\\
2 & uncertainty\\
3 & pixel bitmask\\
4 & wavelength\\
5 & sky\\
6 & sky uncertainty\\
7 & telluric correction spectrum\\
8 & telluric correction uncertainty\\
9 & wavelength coefficients\\
10 & LSF coefficients \\
\enddata
\label{tab:apvisit}
\end{deluxetable}

There will be some bad pixels
in any given spectrum that result from bad pixels on the APOGEE detectors.
Such pixels are flagged in the pixel bitmask that is stored in HDU3
of the \textit{apVisit} files, which encodes the reason why any given pixel
is flagged as unreliable. The bitmask definitions for these are given
at \url{http://www.sdss.org/dr12/algorithms/bitmasks/#APOGEE_PIXMASK}. 
The bitmask also encodes locations where data may be more suspect, e.g.,
pixels in the strong persistence regions of the detectors, and/or
pixels in the location of the Littrow ghost (see \citealt{Wilson2015}).

While correction is made for telluric absorption and for sky emission,
these corrections (especially for sky emission) are generally imperfect.
As a result, users can expect to see some residual sky emission in the
visit spectra. Locations of strong sky emission and telluric absorption
are flagged in the pixel bitmask.

\subsection{Combined spectra}

FITS image files with combined (across multiple visits) spectra of each
star observed with APOGEE are saved in \textit{apStar-VERS-APOGEE\_ID.fits} files.
Here, VERS represents the reduction version (again, r5 for the DR12 release)
and APOGEE\_ID represents an object identifier, which is usually the 2MASS
object name with a '2M' prefix; for the few objects that are not in the 2MASS
catalogs, a similar style APOGEE\_ID was created from the coordinates, and a
'AP' prefix was prepended.

\begin{deluxetable}{ll}[h]
\tablecaption{Contents of \textit{apStar} files}
\tablehead{
\colhead{HDU}&\colhead{contents}
}
\startdata
0 & main header with target, RV, version information\\
1 & resampled spectra \\
2 & uncertainties in spectra \\
3 & pixel mask for spectra \\
4 & sky spectrum \\
5 & sky uncertainty \\
6 & telluric spectrum \\
7 & telluric uncertainty \\
8 & LSF coefficients \\
9 & RV information, including template \\
  & cross-correlation and autocorrelation\\
\enddata
\label{tab:apstar}
\end{deluxetable}

The \textit{apStar} files represent the main survey output spectra. These FITS 
files have multiple HDUs: they include the combined spectra and individual 
resampled visit spectra, uncertainty arrays corresponding to these, a pixel
mask array that flags pixels with various conditions, a wavelength
array, and other information, as summarized in Table \ref{tab:apstar}.

The combined spectrum is included along with the individual visit spectra
shifted to rest wavelength in HDU1. This is a 2D image with wavelength
running along the column axis, and with NVISITS+2 rows that have two
versions of the combined spectra in the first two rows, followed by 
NVISITS rows of resampled individual-visit spectra. The two combinations
are generally very similar; they differ in the weighting scheme used to
do the combination (see \citealt{Nidever2015}). 
The ASPCAP analysis of these spectra uses the first
version of the combined spectra.  Bad pixels in any of the individual
visit spectra are discarded during the combination.

All spectra are put on a common wavelength scale with a fixed dispersion
in log $\lambda$: \textit{apStar} spectra all have 8575 pixels with $\log \lambda_i
= 4.179 + 6.\times 10^{-6} i$, where $i$ is the pixel number (with the first
pixel having $i=0$). This corresponds to a pixel sampling of 4.145 km/s and
provides roughly 3 pixels per resolution element.

Because of the resampling, the uncertainties
are correlated between adjacent pixels; the
covariance information is not currently being tracked. If data with
non-correlated errors are required, users are advised to use the \textit{apVisit}
files, but these have non-constant dispersion and, even here, there will
be small covariances in the dither-combined images because the dithers are
not perfect.

Because spectra are obtained with visits taken over a period of time, telluric
absorption and sky emission features generally do not fall at the same
rest wavelength in the stellar spectra because of the changing barycentric
correction. This means that systematic errors in the correction of these
features are generally reduced compared to the individual visit spectra,
but are spread over a larger region of the stellar spectra. The pixel
bitmask in HDU3 of the \textit{apStar} files track pixels that were affected in
the individual visit spectra.

\subsection{ASPCAP spectra and best fits}

The \textit{aspcapStar} files are FITS image files that give the continuum normalized
spectra from which parameters and abundances are derived. These are given
on the \textit{apStar} wavelength scale, but only include the specific wavelength
ranges that are matched to the synthetic grids. Multiple extensions give
the best-matching synthetic spectra from which the stellar parameters
are adopted (see Table \ref{tab:aspcapstar}).

\begin{deluxetable}{ll}[h]
\tablecaption{Contents of aspcapStar files}
\tablehead{
\colhead{HDU}&\colhead{contents}
}
\startdata
0 & Normalized APOGEE spectrum\\
1 & Uncertainties in normalized spectrum\\
2 & Best library match as derived by ASPCAP/FERRE\\
3 & FITS table with parameters of best fit spectrum\\
\enddata
\label{tab:aspcapstar}
\end{deluxetable}

\subsection{ASPCAP field results }

The \textit{aspcapField} files are FITS tables for each field that include the derived
ASPCAP parameters and abundances for all stars in the field. A second
extension includes a FITS table with all of the normalized spectra and
the best-matching synthetic spectra for all stars in the field. Because
these files include spectra for hundreds of stars, they are moderately
large (see Table \ref{tab:aspcapfield}).

\begin{deluxetable}{ll}[t]
\tablecaption{Contents of aspcapField files}
\tablehead{
\colhead{HDU}&\colhead{contents}
}
\startdata
0 & FITS table with parameters and abundances for all \\
  & \ \ objects in field\\
1 & FITS table with normalized spectra, uncertainty array, \\
  & \ \ and best-fit spectra\\
2 & FITS table giving the order of the parameters and \\
  & \ \ abundances in the arrays of HDU0\\
\enddata
\label{tab:aspcapfield}
\end{deluxetable}

\section{Conclusion}
\label{sect:conclusion}

We have described the data from the SDSS-III/APOGEE survey
as released in the Data Release 12 \citep{DR12}. 

Stellar parameters and abundances have been calibrated and compared with
the literature using observations of stellar clusters and of stars with
previously measured abundances. 
The internal precision of the APOGEE
abundances is typcially 0.05-0.1 dex, judging from the internal scatter
of abundances within clusters.  
External accuracy of the abundances is challenging to
assess, but may be good only to 0.1-0.2 dex. Stellar parameters and
abundances for cooler stars (\teff $<$4000) may be more uncertain.

The APOGEE data set is rich and we expect
that it will enable significant developments in Milky Way science.
We have described the methods by which APOGEE data can be accessed,
and attempt to present information that will provide information about
its reliability and limitations. 

The automatic determination of stellar parameters and abundances presents
a challenging problem, and we do not claim to be currently doing the
most optimal job. Work is ongoing with a number of issues, including:

\begin{itemize}
  \item implementation of a correction for persistence in the SDSS-III/APOGEE data;
  \item improvement of sky emission correction;
  \item improvement of radial velocities, especially for hotter stars and spectra
with lower S/N;
  \item continued revision of line lists;
  \item incorporation of rotation where needed;
  \item extension to cooler temps using the MARCS/Turbospectrum code 
(see \citealt{Zamora2015};
  \item investigation of individual element grids, rather than 
using the element family grids to solve for individual element abundances;
  \item investigation of correlated errors;
  \item improvments in handling of non-detections/upper limits;
  \item investigation of issues with derivation of surface gravity.
\end{itemize}

Nevertheless, the DR12 results have proven to be useful and of a quality 
to conduct a wide variety of science.

As the study of Milky Way stars continues with SDSS-IV/APOGEE-2, there will
be future data releases, and we anticipate that at least some of these will
include improved re-analysis of SDSS-III/APOGEE data.

Funding for SDSS-III has been provided by the Alfred P. Sloan Foundation,
the Participating Institutions, the National Science Foundation, and
the U.S. Department of Energy Office of Science. The SDSS-III web site
is \url{http://www.sdss3.org/}.

SDSS-III is managed by the Astrophysical Research Consortium for the
Participating Institutions of the SDSS-III Collaboration including the
University of Arizona, the Brazilian Participation Group, Brookhaven
National Laboratory, Carnegie Mellon University, University of Florida,
the French Participation Group, the German Participation Group, Harvard
University, the Instituto de Astrofisica de Canarias, the Michigan
State/Notre Dame/JINA Participation Group, Johns Hopkins University,
Lawrence Berkeley National Laboratory, Max Planck Institute for
Astrophysics, Max Planck Institute for Extraterrestrial Physics, New
Mexico State University, New York University, Ohio State University,
Pennsylvania State University, University of Portsmouth, Princeton
University, the Spanish Participation Group, University of Tokyo,
University of Utah, Vanderbilt University, University of Virginia,
University of Washington, and Yale University.  

JAH, SRM, and VVS acknowledge support for this research from the National 
Science Foundation (AST-1109178). JAJ and MP also acknowledge support from the
National Science Foundation (AST-1211673).

TCB acknowledges partial support for this work from grants PHY 08-22648;
Physics Frontier Center/Joint Institute or Nuclear Astrophysics (JINA),
and PHY 14-30152; Physics Frontier Center/JINA Center for the Evolution
of the Elements (JINA-CEE), awarded by the US National Science Foundation.

\bibliographystyle{apj}

\bibliography{ref}

\end{document}